\documentclass[review]{elsarticle}

\usepackage{graphicx,hyperref,url,xcolor,amssymb,amsmath,ctable,booktabs,caption,hhline,epsfig,flexisym,algorithm,algorithmic,float,textcomp,bm}
\usepackage[T1]{fontenc}
\usepackage{amssymb}
\usepackage{enumitem}
\usepackage[autostyle]{csquotes}
\usepackage{amsthm}
\usepackage{multirow}
\usepackage{xcolor}
\PassOptionsToPackage{warn}{textcomp}
\usepackage{textcomp}
\usepackage{booktabs} % For formal tables
\usepackage{url}  %Required
\usepackage{graphicx}  %Required
\usepackage{amssymb}
\usepackage{enumitem}
\usepackage{amsmath}
\usepackage{multirow}
\usepackage{lscape}

\usepackage{xcolor,array,tabulary,mathtools}
\usepackage{url,xcolor,amssymb,amsmath,ctable,caption,hhline,epsfig,flexisym,algorithm,algorithmic,float,bm}
\usepackage[T1]{fontenc}
\usepackage{amssymb}
\usepackage{hyperref}
\usepackage{enumitem}
\usepackage[autostyle]{csquotes}
\usepackage{amsthm}
\usepackage{multirow}
\usepackage{xcolor}
\usepackage{multirow}
\usepackage{xcolor,amssymb,amsmath,ctable,epsfig,float}
\usepackage{subcaption,lipsum,changepage}
\theoremstyle{plain}
\journal{Journal of \LaTeX\ Templates}
\newtheorem*{aman*}{\normalfont \scshape \textbf{\textit{AMAN} Assumption}}
\newtheorem*{theorem1*}{\normalfont \scshape \textbf{Assumption 1}}
\newtheorem*{theorem2*}{\normalfont \scshape \textbf{Assumption 2}}
\newtheorem*{theorem3*}{\normalfont \scshape \textbf{Assumption 3}}
\newcommand{\proposetwo}{\textsf{P3STop}}
\newcommand{\proposethree}{\textsf{P3STop$^{alt}$}}
\newcommand{\proposemix}{\textsf{P3STop$^{mix}$}}
\newcommand{\infpushcr}{\textsf{Inf-Push}}
\newcommand{\propose}{\textsf{{P3S}}}
\newcommand{\proposed}{\textsf{P3S\_1}}
\newcommand{\proposedd}{\textsf{P3S\_2}}
\newcommand{\proposeddd}{\textsf{P3S\_3}}
\newcommand{\etal}{\textit{et al}. }
\usepackage{natbib}

\newcolumntype{C}{>{\centering\arraybackslash}p{5.2em}}
\newcolumntype{K}{>{\centering\arraybackslash}p{3em}}
\begin{document}

\begin{frontmatter}

\title{Click-aware Purchase Prediction with Push at the Top}

\author[uiuc]{Chanyoung Park\fnref{fn1}}
\ead{pcy1302@illinois.edu}
\author[yahoo]{Donghyun Kim}
\ead{donghyun.kim@verizonmedia.com}
\author[naver]{Min-Chul Yang}
\ead{minchul.yang@navercorp.com}
\author[naver]{Jung-Tae Lee}
\ead{jungtae.lee@navercorp.com}
\author[postech]{Hwanjo Yu\corref{cor1}}
\ead{hwanjoyu@postech.ac.kr}
\cortext[cor1]{Corresponding author}
\fntext[fn1]{This work was done during his internship at NAVER.}
\address[uiuc]{Dept. of Computer Science, University of Illinois at Urbana-Champaign, USA}
\address[yahoo]{Yahoo Research, USA}
\address[postech]{Dept. of Computer Science and Engineering, POSTECH, South Korea}
\address[naver]{NAVER Corporation, South Korea}
\begin{abstract}
%Although explicit feedback such as ratings is the most valuable source of information for item recommendation in e-commerce, such feedback is scarce. 
%Consequently, previous studies focused on leveraging the more abundant \textit{implicit} feedback, such as purchase records, for item recommendation. 
%However, 
Eliciting user preferences from purchase records for the task of purchase prediction is challenging because negative feedback is not explicitly observed, and treating all the non-purchased items equally as negative feedback is unrealistic.
%It is challenging to elicit users' preferences from purchase records for purchase prediction because negative feedback is not explicitly observed, and treating all the non-purchased items equally as negative feedback is unrealistic.
In this paper, we present a framework that leverages users' past click records to complement the missing user--item interactions of purchase records, i.e., non-purchased items.
% that reveal users' general interest because users click on numerous items before making purchases.
%In this paper, to complement the missing user--item interactions of purchase records, i.e., non-purchased items, we leverage users' past click records that reveal users' general interest because users click on numerous items before making purchases.
We begin by formulating various model assumptions, each assuming a different order of user preferences among purchased, clicked-but-not-purchased and non-clicked items, to study the usefulness of leveraging click records.
% to complement the missing user--item interactions of purchase records.
We implement the model assumptions under the Bayesian Personalized Ranking model, which maximizes the Area Under the Curve (AUC) for bipartite ranking. 		
%However, we argue that using click records for bipartite ranking needs a meticulously designed model owing to the inherent noisiness of click records compared with purchase records.
However, we argue that using click records for bipartite ranking needs a meticulously designed model owing to the relative unreliableness of click records compared with purchase records.
To address this issue, 
we ultimately propose a novel learning-to-rank method for purchase prediction, called~\proposetwo, that is customized to be robust to relatively unreliable click records by particularly focusing on the accuracy of the top-ranked items.
%we ultimately propose a novel learning-to-rank method,~\proposetwo, 
%that is deliberately designed to be robust to noisy click records.
%Another benefit of~\proposetwo~is that it allows greater focus on the accuracy of the top-ranked items, which is what matters in practice.
% and at the same time focus on the accuracy near the top.
%that is deliberately designed to be robust to noisy click records.
%novel pairwise learning-to-rank method that 1) is robust to noisy click records, and 2) focuses on the accuracy of top-ranked items.
Experimental results on two real-world e-commerce datasets demonstrate that~\proposetwo~considerably outperforms the state-of-the-art implicit feedback--based recommendation methods, especially for the top-ranked items.
\end{abstract}


\end{frontmatter}

%\linenumbers

\section{Introduction}
Implicit feedback, such as purchases and clicks, are easily obtained from system logs, but precisely eliciting users' preferences 
%\vspace{-0.1ex}
from implicit feedback for purchase prediction is challenging because negative feedback is not explicitly observed. 
In this respect, past research has focused on inferring users' negative feedback from missing user--item interactions. Specifically, a uniform weighting scheme~\cite{hu2008collaborative,rendle2009bpr} in which all missing data are treated as negative feedback (i.e., \textit{All Missing As Negative (AMAN)} assumption) has been introduced. However, this assumption is not entirely valid in that 
the reason why items are not observed is uncertain; whether a user does not like them or a user is simply not aware of them.
%we are uncertain whether items are not observed because a user does not like them or simply because the user is not aware of them. 
To cope with the drawback of the \textit{AMAN} assumption, sampling--based approaches such as user-oriented sampling~\cite{pan2008one} or item-popularity-oriented sampling~\cite{he2016fast,rendle2014improving} have been proposed. 
%Users' social network information has also been leveraged, whereby a user is assumed to prefer items that a friend of his has observed to items that have been observed neither by him nor by his friends~\cite{zhao2014leveraging,wang2016social}. 
However, the sampling--based approaches are essentially based on predefined  heuristic weights~\cite{chen2017attentive} that are not guaranteed to always hold in the real data.
%\textcolor{black}{related work is ...}
%and user social network information is not always available in most e-commerce stores.

In this paper, we present a framework that leverages users' past \textit{click records} to complement the missing user--item interactions of purchase records, i.e., non-purchased items, aiming at \textit{purchase prediction}. Precisely, we leverage users' {past click records} in conjunction with their purchase records, both of which are easily collected by e-commerce stores. Intuitively, click records reveal users' general interest because users click on numerous items before making purchases. 
Hence, we expect that users' click records will complement the missing user--item interactions of purchase records in a more data-driven manner compared with previous uniform weighting scheme or sampling--based approaches.
%It is worth noting that such click records have been used for various tasks such as click-through rate (CTR) prediction in online advertising~\cite{mcmahan2013ad,zhu2010novel,zhang2014sequential} and Twitter~\cite{li2015click}, repeat-buyer prediction~\cite{liu2016repeat}, conversion response prediction in display advertising~\cite{li2015predicting}, and session based click prediction or recommendation~\cite{hidasi2015session}. 
%However, not much effort has been devoted to purchase prediction\footnote{Since the goal of recommender systems in the e-commerce domain is to recommend items that are likely to be purchased by users in the future, the prediction of users' purchases can be cast as the task of item recommendation. Therefore, we use these terms, i.e., ``purchase prediction'' and ``item recommendation'' interchangeably hereafter.}, especially to leveraging click records to complement the missing user--item interactions of purchase records.

By making use of click records, we begin by formulating various model assumptions regarding the order of user preferences among the missing user--item interactions of purchase records, i.e., non-purchased items, which can be split into two disjoint sets; \textit{clicked-but-not-purchased} items and \textit{non-clicked} items.
%To this end, by using click records, we 1) split the missing interactions of purchase records, i.e., non-purchased items, into two disjoint sets, i.e., \textit{clicked-but-not-purchased} items and \textit{non-clicked} items, 2) formulate various model assumptions regarding the order of user preferences among these itemsets, and 3) 
%perform experiments to find valid model assumptions. Indeed, 
We empirically demonstrate that a model assumption in which users are assumed to prefer \textit{purchased} (\textsf{P}) items to \textit{clicked-but-not-purchased} (\textsf{CBNP}) items to \textit{non-clicked} (\textsf{NC}) items, is beneficial for purchase prediction when implemented under the Bayesian Personalized Ranking (BPR) model~\cite{rendle2009bpr}, which is a pairwise bipartite ranking model that maximizes the AUC metric. 
To be precise, we make three different positive$-$negative pairs over three disjoint itemsets, i.e., \textsf{P}$-$\textsf{CBNP}, \textsf{CBNP}$-$\textsf{NC} and \textsf{P}$-$\textsf{NC}, and learn a ranking function that is expected to establish a total order in which positive instances precede negative ones in each positive$-$negative pair of itemsets, which is equivalent to maximizing the AUC.
%jointly maximize the AUC over these three pairs.

%However, directly adopting the bipartite ranking model for incorporating click records suffers two limitations.
%However, as the number of click records greatly exceeds that of purchase records, the bipartite ranking model, such as BPR, can be dominated by the click records. To make the matter worse, click records are inherently noisier than purchase records in practice; a user may accidentally click on wrong items or may click on items to see more details and end up not liking it, whereas a user is more confident with purchased items. 
However, \textit{clicks are {weaker} signal of user preference than purchases} in practice. That is, a user may accidentally click on wrong items or may click on items to see more details and end up not liking it, whereas a user is more confident with purchased items. This indicates that clicks are relatively less reliable than purchases in terms of user preference contained therein. To make the matter worse, the number of click records greatly exceeds that of purchase records, implying that the bipartite ranking model such as BPR can be dominated by the relatively unreliable click records.
Therefore, naively incorporating click records for the bipartite ranking can be detrimental to the performance of recommendation\footnote{Since the goal of recommender systems in e-commerce is to recommend items that are likely to be purchased by users, the purchase prediction can be cast as the task of item recommendation. Therefore, we use the terms, i.e., ``purchase prediction'' and ``recommendation'' interchangeably throughout this paper.}, and the model should be meticulously designed to properly harness the click records for purchase prediction under bipartite ranking.
%Another drawback of adopting the bipartite ranking scheme, or equivalently optimizing the AUC metric, is that a mistake in the higher part of the recommendation list equally penalized with one in the lower part, which does not allow a particular focus on the accuracy of the top-ranked items. However, we argue that the \textit{recommendation accuracy of the top-ranked items is especially important} as they get much more attention by users in practice~\cite{top}, and therefore should not be overlooked.
%\vspace{-0.2ex}
To this end, we propose a novel learning-to-rank method for purchase prediction, called \proposetwo, that is customized to be robust to relatively unreliable click records.
%, and at the same time focus on the accuracy near the top.
%especially tailored to incorporating the noisy click records.
%designed to be robust to noisy click records.
%To be precise,
%To address this issue, 
%we build on our valid model assumption, and propose a novel \textit{top-N focused} learning-to-rank method that is deliberately designed to be robust to noisy click records.
%propose a novel pairwise learning-to-rank method that 1) is robust to noisy click records, and 2) focuses on the accuracy of the top-ranked items. 
%We argue that the \textit{recommendation accuracy of the top-ranked items is especially important} as they get much more attention by users in practice~\cite{top}. 
%However, as the bipartite ranking model penalizes a mistake in the higher part of the recommendation list equally with one in the lower part, it does not allow a particular focus on the accuracy of the top-ranked items.
%, while users are mainly interested in top-ranked items~\cite{top}.
%\footnote{http://trends.e-strategyblog.com/2014/11/19/amazon-com-search-click-through-rates-by-position/22261}
More precisely, \textit{\proposetwo~{minimizes the number of ``negative'' items ranked above the last-ranked ``positive'' item}}. As a concrete example, consider the following \textbf{Toy Example} in which we illustrate the push/pull mechanism of modeling the pairwise relationship between \textsf{P} (``positive'') items (in blue), and \textsf{CBNP} (``negative'') items (in green).

\begin{figure}[t]
	\centering
	\includegraphics[width=0.8\textwidth]{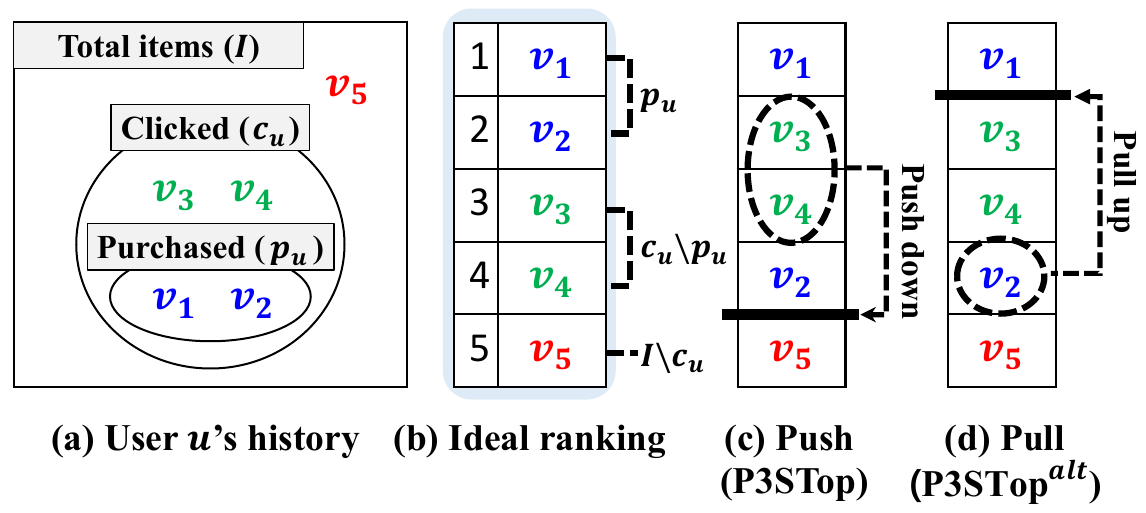}
	%	\vspace{-2.5ex}
	\caption{A toy example of the push/pull mechanism. }
	%	\vspace{-4ex}
	\label{fig:bound}
\end{figure}

\smallskip
\begin{adjustwidth}{0.18cm}{0.18cm}
	\noindent\textbf{\underline{Toy Example.}}	Figure~\ref{fig:bound}a shows user $u$'s interaction history with items ($v_1,v_2,\\v_3,v_4$,$v_5$), and the ideal ranking list for user $u$ is displayed in Figure~\ref{fig:bound}b. That is, for user $u$, we want to train our model so that the items are ordered in the following order at the end of the model training:
	%	our model to order items in the following order: 
	{\textsf{P}} items ($\textbf{p}_u$), \textsf{CBNP} items ($\textbf{c}_u \char`\\ \textbf{p}_u$), \textsf{NC} items ($\mathcal{I} \char`\\ \textbf{c}_u$).
	Assuming that items are incorrectly ranked as in Figure~\ref{fig:bound}c during the training process, we aim to push down as many incorrectly ranked \textsf{CBNP} (``negative'') items, i.e., $v_3,v_4$, below the \textit{bound set by the last-ranked \textsf{P} (``positive'') item}, i.e., $v_2$.
	In other words, we push down the relatively unreliable clicked items below the \textit{bound set by a solid purchased item}, which makes our model more robust to unreliable click records.
	%	 as well as allowing greater focus on the accuracy of top-ranked items. 
	%	As an alternative to the push mechanism described in Figure~\ref{fig:bound}c, the pull mechanism shown in Figure~\ref{fig:bound}d differs in the way that the bound is set. 
	An alternative to the push mechanism (Figure~\ref{fig:bound}c) is the pull mechanism (Figure~\ref{fig:bound}d), which differs in the way that the bound is set.
	Precisely, it pulls up the {purchased} items, i.e., $v_2$, above the \textit{bound set by the first-ranked (``negative'') \textsf{CBNP} item}, i.e., $v_3$, as shown in Figure~\ref{fig:bound}d.
	This method is, however, prone to being dominated by unreliable click records, because the \textit{bound is set by the possibly unreliable clicked item}. In Section~\ref{sec:p3stop} and~\ref{sec:alt}, we will describe the rationale behind each case\footnote{While the above push/pull mechanism is applied to the following pairs of itemsets, i.e., ($\textbf{p}_u \leftrightarrow \textbf{c}_u \char`\\ \textbf{p}_u), (\textbf{c}_u \char`\\ \textbf{p}_u \leftrightarrow \mathcal{I} \char`\\ \textbf{c}_u)$, and $(\textbf{p}_u \leftrightarrow \mathcal{I} \char`\\ \textbf{c}_u)$, we display here only the foremost pair for brevity.}.
\end{adjustwidth}

\smallskip
It is important to note that the above push mechanism in Figure~\ref{fig:bound}c enables the model to \textit{particularly focus on the accuracy of the top-ranked items}.
%A corollary benefit of the above procedure is that it \textit{especially focuses on the accuracy of the top-ranked items} by setting the upper bound of ``negative'' items to the last-ranked ``positive'' item. 
More precisely, the upper bound of \textsf{CBNP} items ($v_3,v_4$) is set to the last-ranked \textsf{P} item ($v_2$), which is the item that the user is more confident with than any \textsf{CBNP} item.
%because it was specifically chosen from all the clicked items. 
In this regard, the bound set by a \textsf{P} item ($v_2$) (Figure~\ref{fig:bound}c) should be relatively high and robust compared with the bound set by a clicked item ($v_3$) (Figure~\ref{fig:bound}d). Therefore, pushing down the incorrectly ranked \textsf{CBNP} items below the last-ranked \textsf{P} item allows greater focus on the accuracy of the top-ranked items, because the bound set by the last-ranked \textsf{P} item is high and robust.
%On the other hand, by setting the lower bound to the first-ranked \textit{clicked-but-not-purchased} item ($v_3$), we cannot guarantee that 
%Recall that the BPR framework is designed to optimize the AUC metric in which a mistake in the higher part of the recommendation list equally penalized with one in the lower part, which does not allow a particular focus on the accuracy of the top-ranked items. 
We argue that our proposed method generates more practical recommendation results for users, since the top-ranked items get much more attention by users in practice~\cite{top}. However, only a few recent studies have particularly considered it for the task of recommendation~\cite{christakopoulou2015collaborative,hu2017decoupled,rafailidis2016joint}.
%although the {recommendation accuracy at the top is especially important} as the top-ranked items get much more attention by users in practice~\cite{top}, only a few recent studies have particularly considered it for the task of recommendation~\cite{christakopoulou2015collaborative,hu2017decoupled,rafailidis2016joint}.
%The push/pull mechanism described above applies for all possible pairs of itemsets, i.e., ($\textbf{p}_u \leftrightarrow \textbf{c}_u \char`\\ \textbf{p}_u), (\textbf{c}_u \char`\\ \textbf{p}_u \leftrightarrow \mathcal{I} \char`\\ \textbf{c}_u)$, and $(\textbf{p}_u \leftrightarrow \mathcal{I} \char`\\ \textbf{c}_u)$.
%Experimental results on two e-commerce datasets demonstrate that our proposed methods considerably outperform the state-of-the-art methods in the task of purchase prediction, especially for top-ranked items.

Our main contributions are summarized as follows:
\begin{enumerate}
	\item To complement the missing user--item interactions of purchase records, we formulate various model assumptions regarding the order of user preferences among non-purchased items by taking the click records into account (\textbf{Section~\ref{sec:4}}).
	\item After we find a valid model assumption under the BPR model, we propose~\proposetwo~that is customized to be robust to relatively unreliable click records by particularly focusing on the accuracy of the top-ranked items.
	(\textbf{Section~\ref{sec:4.2}}).
	\item Experimental results on two real-world e-commerce datasets demonstrate that~\proposetwo~considerably outperforms the state-of-the-art implicit feedback--based recommendation methods, especially for the top-ranked items. (\textbf{Section~\ref{sec:5}}).
\end{enumerate}

It is worth noting that click records have been used for various tasks such as click-through rate (CTR) prediction in online advertising~\cite{mcmahan2013ad,zhu2010novel,zhang2014sequential} and Twitter~\cite{li2015click}, user intent prediction~\cite{cheng2017predicting,lo2016understanding}, repeat-buyer prediction~\cite{liu2016repeat}, conversion response prediction in display advertising~\cite{li2015predicting}, and session--based click prediction~\cite{hidasi2015session}. However, not much effort has been devoted to purchase prediction, and to the best of our knowledge, our work is the first to propose a framework that leverages click records to complement the missing user--item interactions of purchase records.

\begin{table}[]
	\centering
	\captionsetup{font=normal}
	\caption{Notation}
	\label{notation}
	\begin{tabular}{l|l}
		\specialrule{.1em}{.1em}{.1em} 
		Symbol & \multicolumn{1}{c}{Description} \\ \specialrule{.1em}{.1em}{.1em}
		$\mathcal{U}, \mathcal{I}$& Set of Users, Set of Items                                 \\ 		
		$n, m$& Number of users and items                                 \\ 
		$\textbf{P}\in\mathbb{R}^{n\times m}$& User-Item Purchase matrix \\ 
		$\textbf{C}\in\mathbb{R}^{n\times m}$& User-Item Click matrix \\ 
		$\textbf{p}_u$& Items purchased by user $u$ \\ 
		$\textbf{c}_u$& Items clicked by user $u$ \\ 		
		$K$& Number of latent dimensions                                 \\ 
		$\bm{\alpha \in \mathbb{R}}^{n\times K}$& User latent matrix                                 \\ 
		$\bm{\beta \in \mathbb{R}}^{m\times K}$& Item latent matrix                                 \\ 
		$\bm{\gamma \in \mathbb{R}}^{m}$& Item bias                                 \\ 
		${\lambda}$& The strength of the model regularization                                 \\ 
		${\eta}$& Learning rate \\\specialrule{.1em}{.1em}{.1em}
	\end{tabular}
\end{table}

%\vspace{-1ex}
\section{Problem Statement}
%\vspace{-.1ex}
We first introduce notations used throughout this paper (Table~\ref{notation}). 
Let $\mathcal{U}$ and $\mathcal{I}$ be the set of users and items, respectively, and we have $n$ users and $m$ items.
The purchase records of users in $\mathcal{U}$ on items in $\mathcal{I}$ are represented by the purchase matrix $\textbf{P} = {[{p_{ui}}]_{n \times m}}$, where ${p_{ui}=1}$ if user $u$ purchased item $i$, and 0 otherwise. Likewise, the click records of users in $\mathcal{U}$ on items in $\mathcal{I}$ are represented by the click matrix $\textbf{C} = {[{c_{ui}}]_{n \times m}}$, where ${c_{ui}=1}$ if user $u$ clicked item $i$, and 0 otherwise; counts are ignored in this work.
$\textbf{p}_u$ and $\textbf{c}_u$ denote the sets of items purchased and clicked by user $u$, respectively. We formally define our problem in this paper as follows:

\smallskip
\noindent\textbf{\textit{Problem Definition}}\mbox{} \\
\noindent\textbf{Given: }The purchase matrix ${\textbf{P}}$ and click matrix ${\textbf{C}}$, 

\noindent\textbf{Goal: }To recommend items $i\in \mathcal{I} \textbackslash (\textbf{p}_u \cup \textbf{c}_u)$ to each user $u \in \mathcal{U}$; among items that the user has not previously interacted with (neither purchased nor clicked). 
%\section{Method}
%In this section, we first describe our approach to dealing with missing user--item interactions of purchase records.
%%aiming at predicting users' future purchases in e-commerce. 
%We explain our model assumptions regarding the fine-grained order of user preferences among non-purchased items (\textbf{Section~\ref{sec:4.1}}). 
%Next, we describe how our model assumptions are implemented under the BPR framework (\textbf{Section~\ref{sec:4.1.1}}).
%Then, we discuss two limitations of incorporating click records under the BPR framework
%%a challenge of incorporating noisy click records into the model, and a limitation of the AUC metric for bipartite ranking, which BPR is designed to maximize 
%(\textbf{Section~\ref{sec:4.1.2}}).
%Finally, we describe our proposed method, called~\proposetwo, to overcome these limitations (\textbf{Section~\ref{sec:4.2}}).

%\vspace{-0.5ex}
\section{Ordering User Preferences among Non-purchased Items}
\label{sec:4}

In this section, we describe our framework that leverages click records to complement the missing user--item interactions of purchase records. i.e., non-purchased items.
%aiming at predicting users' future purchases in e-commerce. 
We begin by explaining our model assumptions regarding the order of user preferences among non-purchased items (\textbf{Section~\ref{sec:4.1}}). 
Next, we describe how our model assumptions are implemented under the BPR model (\textbf{Section~\ref{sec:4.1.1}}).
Then, we discuss two shortcomings of naively incorporating click records under the BPR model
%a challenge of incorporating noisy click records into the model, and a limitation of the AUC metric for bipartite ranking, which BPR is designed to maximize 
(\textbf{Section~\ref{sec:4.1.2}}).

%\vspace{-1ex}
\subsection{Defining the Model Assumptions}
\label{sec:4.1}
Recall the \textit{AMAN} assumption made by previous pairwise learning-to-rank methods~\cite{he2016vbpr,pan2013gbpr,rendle2009bpr}. 

\begin{aman*}
	%	\vspace{-1ex}
	\leftskip=0.2cm
	\normalfont	We assume that a user prefers \textit{purchased} items to \textit{non-purchased} items.
	\rightskip=0.2cm
	\begin{equation}
		\label{AMAN}
		i \succ_u j, \text{if}~ i \in \textbf{p}_u \wedge  j \in \mathcal{I} \char`\\\textbf{p}_u
	\end{equation}
	Eqn.~\ref{AMAN} implies that user $u$ prefers purchased items $i$ to non-purchased items $j$. However, this assumption is oversimplified in that all non-purchased items are equally considered as negative feedback, whereas in reality some of the non-purchased items attract the user more than the others.
	%\vspace{-1ex}
\end{aman*}
%For each user $u$ and his purchased set of items $\textbf{p}_u$, the following assumption is made:
%\begin{equation}
%\label{AMAN}
%i \succ_u j, \text{if}~ i \in \textbf{p}_u \wedge  j \in \mathcal{I} \char`\\\textbf{p}_u
%\end{equation}
To overcome the above limitation of the \textit{AMAN} assumption, we incorporate users' click records that reveal users' general interest, assuming that users click on numerous items before making purchases. Although the user preference reflected therein is not as strong as in purchase records, we expect that 
%	\textit{click records will help to relieve the AMAN assumption when combined with purchase records.}
\textit{click records will complement the missing user--item interactions of purchase records.}
To this end, given \textit{purchased} items, we split the non-purchased items into two disjoint sets, i.e.,  \textit{clicked-but-not-purchased} items and \textit{non-clicked} items, by using click records, and introduce three different model assumptions regarding the order of user preferences among them. For each user $u$, we assume \textcolor{black}{$\textbf{p}_u \subset \textbf{c}_u \subset \mathcal{I}$}, i.e., all purchased items are selected from clicked items.
\begin{theorem1*}
	%	\vspace{-1.5ex}
	\leftskip=0.2cm
	\normalfont	We assume that a user prefers \textit{purchased} items to \textit{non-clicked} items.
	\rightskip=0.2cm
	\begin{equation}
		\label{assumption1}
		i \succ_u j, \text{if}~  i \in \textbf{p}_u \wedge  j \in \mathcal{I} \char`\\\textbf{c}_u
	\end{equation}
	Instead of regarding non-purchased items as negative feedback as in Eqn.~\ref{AMAN}, this time we regard non-clicked items as negative feedback. This narrows down the candidates for negative feedback, i.e., from $\mathcal{I} \char`\\\textbf{p}_u$ to $\mathcal{I} \char`\\\textbf{c}_u$, which is expected to relieve the \textit{AMAN} assumption.
	%	\vspace{-0.7ex}
\end{theorem1*}

\begin{theorem2*}
	%	\vspace{-0.7ex}
	\leftskip=0.2cm
	\normalfont	We assume that a user prefers \textit{purchased} items to \textit{clicked-but-not-purchased} items, \textit{clicked-but-not purchased} items to \textit{non-clicked} items\textcolor{black}{, and \textit{purchased} items to \textit{non-clicked} items.}
	\rightskip=0.2cm
	\begin{equation}
		\label{assumption2}
		i \succ_u j,\; j \succ_u k, \; i \succ_u k, \text{if}~  i \in \textbf{p}_u \wedge  j \in \textbf{c}_u \char`\\ \textbf{p}_u \wedge  k \in \mathcal{I} \char`\\ \textbf{c}_u
	\end{equation}
	We extend \scshape \textbf{Assumption 1} \normalfont by adding another set of items. i.e., {clicked-but-not-purchased} items ($\textbf{c}_u\char`\\\textbf{p}_u$). Eqn.~\ref{assumption2} is based on the assumption that 1) user $u$ is more confident with purchased items ($\textbf{p}_u$) than to clicked-but-not-purchased items ($\textbf{c}_u\char`\\\textbf{p}_u$), because users generally decide to purchase items over many other candidates ($\textbf{c}_u\char`\\\textbf{p}_u$) that reveal users' general interest, which implies that 2) user $u$ prefers clicked-but-not-purchased items to the items that are neither purchased nor clicked ($\mathcal{I} \char`\\\textbf{c}_u$).
	%	\vspace{-0.7ex}
\end{theorem2*}

\begin{theorem3*}
	%	\vspace{-0.7ex}
	\leftskip=0.2cm
	\normalfont	We assume that a user prefers \textit{purchased} items to \textit{clicked-but-not-purchased} items, and \textit{non-clicked} items to \textit{clicked-but-not-purchased} items.
	\rightskip=0.2cm
	\begin{equation}
		\label{assumption3}
		i \succ_u j,\; k \succ_u j, \; i \succ_u k, \text{if}~  i \in \textbf{p}_u \wedge  j \in \textbf{c}_u \char`\\ \textbf{p}_u \wedge  k \in \mathcal{I} \char`\\ \textbf{c}_u
	\end{equation}
	Eqn.~\ref{assumption3} implies that user $u$ dislikes items that are clicked-but-not-purchased ($\textbf{c}_u\char`\\\textbf{p}_u$) more than those that are not clicked at all ($\mathcal{I} \char`\\\textbf{c}_u$). 
	This assumption is also intuitive in the sense that although being aware of clicked-only items ($\textbf{c}_u\char`\\\textbf{p}_u$), the fact that the user still chose not to purchase them implies that the user dislikes them.
	%	\vspace{-0.8ex}
\end{theorem3*}

\subsection{\textbf{Verifying the Model Assumptions}}
\label{sec:4.1.1}
To figure out which of our three model assumptions (Eqn.~\ref{assumption1},\ref{assumption2},\ref{assumption3}) is valid, we implement them under the BPR model~\cite{rendle2009bpr}, and name each of them~\proposed, \proposedd~and~\proposeddd, respectively
(\propose~stands for modeling \underline{p}airwise relationships among \underline{three} disjoint item \underline{s}ets).
We only present here the equation for~\proposedd, which is based on \scshape \textbf{Assumption 2}\normalfont. The equations for~\proposed~and~\proposeddd\ are similarly formulated and hence omitted.
%Given parameters $\Theta$ for training, we maximize the following likelihood function for~\proposedd:
For each user $u$, we maximize the following loss function:
%\begin{equation}
%\label{likelihood_P3S2}
%\begin{split}
%&\mathcal{J}_{\proposedd}(\Theta)=\prod\limits_{u\in\mathcal{U}} ( \prod\limits_{i\in \textbf{p}_u}\prod\limits_{j\in \textbf{c}_u \char`\\ \textbf{p}_u} \Pr[i \succ_u j] \\
%&\prod_{j\in \textbf{c}_u \char`\\ \textbf{p}_u}\prod_{k\in \mathcal{I} \char`\\ \textbf{c}_u} \Pr[j \succ_u k] \prod_{i\in \textbf{p}_u}\prod_{k\in \mathcal{I} \char`\\ \textbf{c}_u} \Pr[i \succ_u k])
%\end{split}
%\end{equation}
\begin{equation}
	\label{likelihood_P3S2}
	\begin{split}
		&{\mathcal{L}_{\proposedd}}(u) = \sum\limits_{i \in {\textbf{p}_u}} {\sum\limits_{j \in {\textbf{c}_u}\backslash {\textbf{p}_u}} {\ln \sigma ({{\hat x}_{uij}})} }  \\
		& + \sum\limits_{j \in {\textbf{c}_u}\backslash {\textbf{p}_u}} {\sum\limits_{k \in I\backslash {\textbf{c}_u}} {\ln \sigma ({{\hat x}_{ujk}})} }  + \sum\limits_{i \in {\textbf{p}_u}} {\sum\limits_{k \in I\backslash {\textbf{c}_u}} {\ln \sigma ({{\hat x}_{uik}})} } 
	\end{split}
\end{equation}
%where $\Pr[i \succ_u j]= \sigma(\hat x_{ui} - \hat x_{uj})$ denotes the probability that user $u$ prefers item $i$ to item $j$, which is approximated by a sigmoid function of the form $\sigma(x)={1}/(1+e^{-x})$~\cite{rendle2009bpr},
%and $\hat x_{ui} = \alpha_u^T\beta_i + \gamma_i$ denotes the predicted preference of user $u$ on item $i$ computed by matrix factorization (MF); $\alpha_u\in\mathbb{R}^K$ and $\beta_i\in\mathbb{R}^K$ represent the $K$-dimensional latent factors for user $u$ and item $i$, respectively, and $\gamma_i\in\mathbb{R}$ denotes the item bias term for item $i$. 
where ${\hat x}_{uik}={\hat x}_{ui} - {\hat x}_{uk}$, and $\hat x_{ui} = \alpha_u^T\beta_i + \gamma_i$ denotes the predicted preference of user $u$ on item $i$ computed by matrix factorization (MF); $\alpha_u\in\mathbb{R}^K$ and $\beta_i\in\mathbb{R}^K$ represent the $K$-dimensional latent factors for user $u$ and item $i$, respectively, and $\gamma_i\in\mathbb{R}$ denotes the item bias term for item $i$.  $\sigma ({{\hat x}_{uij}})$ denotes the probability that user $u$ prefers item $i$ to item $j$~\cite{rendle2009bpr}, which is approximated by a sigmoid function $\sigma(\cdot)$.
For more details of the optimization process, refer to the original paper~\cite{rendle2009bpr} that proposed the BPR model. We later show in our experiments (Table~\ref{tab:P3S}) that~\proposedd~outperforms~\proposed~and~\proposeddd, which implies that \scshape \textbf{Assumption 2}\normalfont~is the most valid model assumption.

Note that other scoring functions such as neural network (NN)--based functions
\textcolor{black}{
~\cite{he2017neural,hidasi2015session,landin2019prin}} can also be applied to our framework by simply replacing MF.
However, as our focus is to propose a ``framework'' that can properly utilize click records for purchase prediction rather than to prove the superiority of NN over MF, we conduct experiments with MF as our scoring function in this paper.

%\vspace{-0.8ex}
\subsection{\textbf{Discussion: Shortcomings of~\proposedd}}
\label{sec:4.1.2}
Although~\proposedd~is shown to be beneficial for purchase prediction when implemented under the BPR model,
it has two shortcomings. 
%The first shortcoming is caused by the relatively large amount of click records, which is even noisier in nature compared with purchase records. Unlike purchases, clicks can occur even without a user's intent to purchase; a user may accidentally click on wrong items or click on items out of simple curiosity, whereas a user is more confident with purchased items.
The first shortcoming is caused by the relative unreliableness of click records, whose amount even greatly exceeds that of purchase records. Unlike purchases, clicks can occur even without a user's intent to purchase; a user may accidentally click on wrong items or click on items out of simple curiosity, whereas a user is more confident with purchased items.
That is to say, the click records are more likely to be irrelevant to user preferences than the purchase records.
Therefore, we argue that relying too much on the relatively unreliable click records would be detrimental to the performance of recommendation.
However, since BPR was developed for bipartite ranking in which every possible positive--negative instance pair is taken into account, the model can be easily dominated by the relatively unreliable click records as their amount greatly exceeds that of purchase records.
The second shortcoming is caused by the objective of the BPR model. Although users are mainly interested in top-ranked items~\cite{top}, BPR maximizes the AUC, which gives an equal weight to each training instance regardless of its position in the list. In other words, a mistake in the higher part of the recommendation list is equally penalized with one in the lower part, implying that optimizing the AUC does not allow a particular focus on the accuracy of the top-ranked items.
Therefore, we propose a novel method that \textit{simultaneously addresses} the above shortcomings.

%\vspace{-0.5ex}
\section{The Proposed Method:~\proposetwo}
\label{sec:4.2}
%Here, we describe our proposed method, called~\proposetwo, which is robust to noisy click records, and at the same time focus on the accuracy near the top.
Here, we describe our novel learning-to-rank method,~\proposetwo, that is customized to be \textit{robust to relatively unreliable click records} by particularly \textit{focusing on the accuracy of the top-ranked items}.
%1) is robust to noisy click records, and 2) optimizes the ranking accuracy especially near the top of the ranked list of recommendations. 
Since~\proposedd, which is based on \scshape \textbf{Assumption 2}\normalfont, turned out to be the most valid model (Table~\ref{tab:P3S}), we adopt it as the underlying assumption of our proposed method,~\proposetwo, hereinafter. 
Note that under \scshape \textbf{Assumption 2}\normalfont, 
{purchased} (\textsf{P}) items and {non-clicked} (\textsf{NC}) items are always regarded as positive and negative items, respectively.
In contrast, {clicked-but-not-purchased} (\textsf{CBNP}) items can be considered as either positive or negative items, depending on which pair of itemsets we are interested in.
That is, \textsf{CBNP} items are considered as negative items when compared with \textsf{P} items, and are considered as positive items when compared with \textsf{NC} items.

%Note that under \scshape \textbf{Assumption 2}\normalfont, \textit{purchased} items and \textit{non-clicked} items are regarded respectively as positive and negative items, whereas \textit{clicked-but-not-purchased} items can be considered as either positive items or negative items, depending on which pair of itemsets we are interested in.

%\smallskip
%\noindent\textbf{Model Formulation}
%\vspace{-0.9ex}
\subsection{\textbf{Model Formulation}}
%\vspace{-0.2ex}
\label{sec:p3stop}
For each user $u$, we compute the sum of the number of 1) \textsf{CBNP} items ranked above the least relevant \textsf{P} item, 2) \textsf{NC} items ranked above the least relevant \textsf{CBNP} item, and 3) \textsf{NC} items ranked above the least relevant \textsf{P} item, and minimize the sum as following:
\textcolor{black}{
\begin{equation}
	\label{loss}
	\small
	\begin{split}
		\mathcal{L}_{\proposetwo}(u) 
		&= \frac{1}{{|{\textbf{c}_u}\char`\\{\textbf{p}_u}|}}\sum\limits_{j \in {\textbf{c}_u}\char`\\{\textbf{p}_u}} \mathbb{I}{\left[( \mathop {\min }\limits_{i \in {\textbf{p}_u}} {\hat{x}_{ui}}) \le {\hat{x}_{uj}} \right]} \\
		&+ \frac{1}{{|I\char`\\{\textbf{c}_u}|}}\sum\limits_{k \in I\char`\\{\textbf{c}_u}} \mathbb{I}{\left[ (\mathop {\min }\limits_{j \in {\textbf{c}_u}\char`\\{\textbf{p}_u}} {\hat{x}_{uj}}) \le {\hat{x}_{uk}} \right]} \\
		& + \frac{1}{{|I\char`\\{\textbf{c}_u}|}}\sum\limits_{k \in I\char`\\{\textbf{c}_u}} \mathbb{I}
		\left[ (\mathop {\min }\limits_{i \in {\textbf{p}_u}} {\hat{x}_{ui}}) \le {\hat{x}_{uk}} \right] \\
		& = \frac{1}{{|{\textbf{c}_u}\char`\\{\textbf{p}_u}|}}\sum\limits_{j \in {\textbf{c}_u}\char`\\{\textbf{p}_u}} {max\left[0,  1-((\mathop {\min }\limits_{i \in {\textbf{p}_u}} {\hat{x}_{ui}}) - {\hat{x}_{uj}}) \right]} \\
		& + \frac{1}{{|I\char`\\{\textbf{c}_u}|}}\sum\limits_{k \in I\char`\\{\textbf{c}_u}} max{\left[0, 1-((\mathop {\min }\limits_{j \in {\textbf{c}_u}\char`\\{\textbf{p}_u}} {\hat{x}_{uj}}) - {\hat{x}_{uk}}) \right]} \\
		& + \frac{1}{{|I\char`\\{\textbf{c}_u}|}}\sum\limits_{k \in I\char`\\{\textbf{c}_u}} max
		\left[0, 1-((\mathop {\min }\limits_{i \in {\textbf{p}_u}} {\hat{x}_{ui}}) - {\hat{x}_{uk}}) \right]
	\end{split}
	\raisetag{40pt}
\end{equation}
}
where $\hat x_{ui} = \alpha_u^T\beta_i$\footnote{The incorporation of the item bias term ($\gamma_i$) as in Eqn.~\ref{likelihood_P3S2} did not result in the performance improvement, hence excluded.} and $\mathbb{I}[\cdot]$ is the indicator function \textcolor{black}{that returns 1 if the argument is true, otherwise 0}.
%Note that Eqn.~\ref{loss} gives more weight to positive items than to negative items, which makes our model robust to noisy negative items.
%Take the first term as an example, where we set the upper bound score of the clicked-but-not-purchased items ($\textbf{c}_u\char`\\\textbf{p}_u$) to the score of the last-ranked purchased item ($\textbf{p}_u$) (Figure~\ref{fig:bound}c). Although the last-ranked purchased item has the lowest score among all purchased items, its score should be definitely higher than that of any clicked-but-not-purchased items because it was selected from among all clicked items. Consequently, the upper bound will be high enough so that we obtain relevant items near the top by pushing down negative items below it.
%In summary, by minimizing $\mathcal{L}_{\proposetwo}(u)$ for each user, we aim to put as many negative items below positive items as possible; this results in a high accuracy, especially for the top-ranked items.
Note that in Eqn.~\ref{loss} the bound is set with respect to positive items and thus more emphasis is placed on positive items than on negative items, making our model robust to relatively unreliable negative items.
For example, consider the first term, where for user $u$, we set the upper bound of the \textsf{CBNP} items $j$ ($\in\textbf{c}_u\char`\\\textbf{p}_u$) to the score of the last-ranked \textsf{P} item $({\min _{i \in {{\mathbf{p}}_u}}}{{\hat x}_{ui}})$ (Figure~\ref{fig:bound}c). Although the last-ranked \textsf{P} item $i$ has the lowest score among all \textsf{P} items $\mathbf{p}_u$, its score (${\min _{i \in {{\mathbf{p}}_u}}}{{\hat x}_{ui}}$) should be higher than the score ($\hat{x}_{uj}$) of any \textsf{CBNP} item $j$ ($\in {\textbf{c}_u}\char`\\{\textbf{p}_u}$) because it was specifically chosen by user $u$ from all the clicked items (\scshape \textbf{Assumption 2}\normalfont). Consequently, the upper bound set by the last-ranked positive item will be high enough so that we obtain positive items near the top by pushing down relatively unreliable negative items below it.
In summary, by minimizing $\mathcal{L}_{\proposetwo}(u)$ for each user, we aim to put as many unreliable negative items below positive items as possible, which results in high accuracy especially for the top-ranked items.
%As for the optimization of Eqn.~\ref{loss}, since $\mathbb{I}[\cdot]$ is non-convex, we replace it with the hinge loss function $\ell(x) = max(0, 1-x)$, which is a widely used convex surrogate for the indicator function~\cite{rudin2009p}.
As for the optimization of Eqn.~\ref{loss}, since $\mathbb{I}[\cdot]$ is non-convex, making the optimization process difficult because of its discrete nature, we replace it with the hinge loss function $\ell(x) = max(0, 1-x)$, which is a widely used convex surrogate for the indicator function~\cite{rudin2009p}.

\smallskip
\noindent\textbf{Optimization Objective. }
Given the loss function $\mathcal{L}_{\proposetwo}(u)$ for each user $u\in\mathcal{U}$ as in Eqn.~\ref{loss}, the final objective function to minimize is formulated as follows:
\begin{equation}
	%\small
	\label{final_model}
	\mathcal{J}_{\proposetwo}(\Theta) = \frac{1}{{|\mathcal{U}|}}\sum_{u\in \mathcal{U}}\mathcal{L}_{\proposetwo}(u) + \frac{\lambda_{\alpha}}{2} \sum_{u\in \mathcal{U}} ||\alpha_u||_2^2 + \frac{\lambda_{\beta}}{2} \sum_{i\in \mathcal{I}} ||{\beta_i}||_2^2 
\end{equation}
where $\lambda_{\alpha}$ and $\lambda_{\beta}$ are regularization parameters for the user and for the item latent factors, respectively. 
We set $\lambda_{\alpha} = \lambda_{\beta} = \lambda$ to reduce the model complexity.
We adopt the widely used stochastic gradient descent (SGD) method to optimize the objective function in Eqn.~\ref{final_model}. For each user $u$, we first sample a triple $(i,j,k)$ from the training set $\mathcal{O}=\{\mathcal{O}_u | u\in|\mathcal{U}|\}$ where
%	\vspace{-0.5ex}
%	\begin{equation}
$
\mathcal{O}_u=\{(i,j,k)|i \in \textbf{p}_u \wedge  j \in \textbf{c}_u \char`\\ \textbf{p}_u \wedge  k \in \mathcal{I} \char`\\ \textbf{c}_u\}
%	\vspace{-0.5ex}
$
%	\end{equation}
, $\mathcal{O}_u$ denoting the training set for user $u$. We compute the gradient for each parameter in $\Theta$, i.e., $\alpha_u, \beta_i, \beta_j, \beta_k$, and update each of them by using SGD. The gradient for each parameter is computed as follows:
\begin{itemize}
	\item The gradient of $\alpha_u$ for $u\in\mathcal{U}$:
	\begin{equation}
		\label{g_alpha}
		%	\small
		\begin{split}
			&\frac{{\partial L_{\proposetwo}(u)}}{{\partial {\alpha _u}}} = \mathbb{I}[{(\mathop {\min }\limits_{i \in {\textbf{p}_u}} {\hat{x}_{ui}}) - {\hat{x}_{uj}} \le 1]} \left({\beta _j} - \frac{{\partial \mathop {\min }\limits_{i \in {\textbf{p}_u}} {\hat{x}_{ui}}}}{{\partial {\alpha _u}}}\right) \\
			& + \mathbb{I}{[(\mathop {\min }\limits_{j \in {\textbf{c}_u}\char`\\{\textbf{p}_u}} {\hat{x}_{uj}} )- {\hat{x}_{uk}} \le 1]} \left({\beta _k} - \frac{{\partial \mathop {\min }\limits_{j \in {\textbf{c}_u}\char`\\{\textbf{p}_u}} {\hat{x}_{uj}}}}{{\partial {\alpha _u}}}\right) \\
			& + \mathbb{I}{[(\mathop {\min }\limits_{i \in {\textbf{p}_u}} {\hat{x}_{ui}} )- {\hat{x}_{uk}} \le 1]} \left({\beta _k} - \frac{{\partial \mathop {\min }\limits_{i \in {\textbf{p}_u}} {\hat{x}_{ui}}}}{{\partial {\alpha _u}}}\right)
		\end{split}
	\end{equation}
\end{itemize}

\begin{itemize}
	\item The gradient of $\beta_i$ for $i\in\textbf{p}_u$
	\begin{equation}
		\label{g_betai}
		%	\small
		\begin{split}
			\frac{{\partial \mathcal{L}_{\proposetwo}(u)}}{{\partial {\beta _i}}} &=  \mathbb{I}{\left[ {(\mathop {\min }\limits_{i \in {\textbf{p}_u}} {\hat{x}_{ui}}) - {\hat{x}_{uj}} \le 1} \right]} \left( { - \frac{{\partial \mathop {\min }\limits_{i \in {\textbf{p}_u}} {\hat{x}_{ui}}}}{{\partial {\beta _i}}}} \right)\\
			& + \mathbb{I}{\left[ {(\mathop {\min }\limits_{i \in {\textbf{p}_u}} {\hat{x}_{ui}} )- {\hat{x}_{uk}} \le 1} \right]}
			\left( { - \frac{{\partial \mathop {\min }\limits_{i \in {\textbf{p}_u}} {\hat{x}_{ui}}}}{{\partial {\beta _i}}}} \right)
		\end{split}
	\end{equation}
\end{itemize}

\begin{itemize}
	\item The gradient of $\beta_j$ for $j \in \textbf{c}_u \char`\\ \textbf{p}_u$
	\begin{equation}
		\label{g_betaj}
		%	\small
		\begin{split}
			\frac{{\partial \mathcal{L}_{\proposetwo}(u)}}{{\partial {\beta _j}}} &= \mathbb{I}[ {(\mathop {\min }\limits_{i \in {\textbf{p}_u}} {\hat{x}_{ui}}) - {\hat{x}_{uj}} \le 1} ]( {{\alpha _u}} )\\
			& + \mathbb{I} {\left[ {(\mathop {\min }\limits_{j \in {\textbf{c}_u}\char`\\{\textbf{p}_u}} {\hat{x}_{uj}}) - {\hat{x}_{uk}} \le 1} \right]} \left( { - \frac{{\partial \mathop {\min }\limits_{j \in {\textbf{c}_u}\char`\\{\textbf{p}_u}} {\hat{x}_{uj}}}}{{\partial {\beta _j}}}} \right)
		\end{split}
	\end{equation}
\end{itemize}

\begin{itemize}
	\item The gradient of $\beta_k$ for $k \in \mathcal{I} \char`\\ \textbf{c}_u$
	\begin{equation}
		\label{g_betak}
		%	\small
		\begin{split}
			\frac{{\partial \mathcal{L}_{\proposetwo}(u)}}{{\partial {\beta _k}}} &=  \mathbb{I}{[(\mathop {\min }\limits_{j \in {\textbf{c}_u}\char`\\{\textbf{p}_u}} {\hat{x}_{uj}}) - {\hat{x}_{uk}} \le 1]} ({\alpha _u}) \\
			&+ \mathbb{I}{[(\mathop {\min }\limits_{i \in {\textbf{p}_u}} {\hat{x}_{ui}}) - {\hat{x}_{uk}} \le 1]} ({\alpha _u})
		\end{split}
		\raisetag{30pt}
	\end{equation}
\end{itemize}

%Note that the algorithm terminates when the accuracy on the validation set gives the best result.

%Note that the algorithm terminates when the difference in loss between two iterations is less than 0.000001 or when the predetermined number of iterations is reached, which is set to 80\footnote{Every competitor including our method converged before 80 iterations.}.

\textcolor{black}{
Note that the derivatives given by:
\begin{equation}
\begin{dcases}
\frac{{\partial \mathop {\min }\limits_{i \in {X}} {\hat{x}_{ui}}}}{{\partial {\alpha _u}}}=\beta_i\text{, where } \alpha_u^T\beta_i \text{ is the minimum for } i\in X\\
\frac{{\partial \mathop {\min }\limits_{i \in X} {\hat{x}_{ui}}}}{{\partial {\beta _i}}}
= \alpha_u
\end{dcases}
\end{equation}
where $X\in\{\textbf{p}_u, {\textbf{c}_u}\char`\\{\textbf{p}_u}\}$.
}
\subsection{\textbf{Alternative Method:~\proposethree}}
\label{sec:alt}
%\noindent\textbf{Alternative Model. }
As an alternative to our proposed method~\proposetwo, we can consider another method that relies more heavily on click records.
For each user $u$, we compute the sum of the number of 1) \textsf{P} items ranked below the most relevant \textsf{CBNP} item, 2) \textsf{CBNP} items ranked below the most relevant \textsf{NC} item, and 3) \textsf{P} items ranked below the most relevant \textsf{NC} item, and minimize the sum as follows:
\begin{equation}
	\label{height}
	%\small
	\begin{split}
		\mathcal{L}_{\proposethree}(u) &= \frac{1}{{|{\textbf{p}_u}|}}\sum\limits_{i \in {\textbf{p}_u}} \mathbb{I}{\left[ {{\hat{x}_{ui}} \le (\mathop {\max }\limits_{j \in {\textbf{c}_u}\char`\\{\textbf{p}_u}} {\hat{x}_{uj}}}) \right]} \\
		& + \frac{1}{{|{\textbf{c}_u}\char`\\{\textbf{p}_u}|}}\sum\limits_{j \in {\textbf{c}_u}\char`\\{\textbf{p}_u}} \mathbb{I}{\left[ {{\hat{x}_{uj}} \le (\mathop {\max }\limits_{k \in I\char`\\{\textbf{c}_u}} {\hat{x}_{uk}}}) \right]} \\
		&+ \frac{1}{{|{\textbf{p}_u}|}}\sum\limits_{i \in {\textbf{p}_u}} \mathbb{I}{\left[ {{\hat{x}_{ui}} \le (\mathop {\max }\limits_{k \in I\char`\\{\textbf{c}_u}} {\hat{x}_{uk}}}) \right]}
	\end{split}
	%\raisetag{40pt}
\end{equation}
This method, named~\proposethree, is distinguished from~\proposetwo~in that~\proposethree~resorts to the negative items to set the bound. To be precise, it sets the \textit{lower bound of the positive items to the score of the top-ranked negative item}; as opposed to~\proposetwo~that sets the \textit{upper bound of negative items to the last-ranked positive item}. Here, we want the lower bound to be high enough so that pulling up the positive items above it is meaningful (Figure~\ref{fig:bound}d). 
However, the negative items always include relatively unreliable click records in this case, and thus the lower bound set by the negative items is not guaranteed to be sufficiently high and robust; in contrast to high and robust bound of~\proposetwo~set by positive items.
This implies that pulling up positive items above relatively low bound would not yield high accuracy at the top.
%However, if the negative items are noisy as in our case, the lower bound score may not be high enough, so pulling up positive items above it would not yield high accuracy at the top.
We later present the performance of~\proposethree~in the experiments (Table~\ref{tab:P3SPush}) to show the unreliableness of click records compared with purchase records.
We note that~\proposethree~is an enhanced version of~\infpushcr~\cite{christakopoulou2015collaborative}, whose underlying \textit{AMAN} assumption is replaced with \scshape \textbf{Assumption 2}\normalfont~\cite{li2014top}.

\smallskip
\noindent\textbf{Complexity Analysis.}
Another benefit of~\proposetwo~is the improved time complexity compared with previous pairwise methods developed for bipartite ranking, such as BPR and~\propose. More precisely, the time complexity of evaluating ${\proposedd}$ is $O(|\textbf{p}_u||\textbf{c}_u \char`\\ \textbf{p}_u| + |\textbf{c}_u \char`\\ \textbf{p}_u||\mathcal{I} \char`\\ \textbf{c}_u| + |\textbf{p}_u||\mathcal{I} \char`\\ \textbf{c}_u|)=O(m^2)$ whereas that for ${\proposetwo}$ is $O(2\times(|\textbf{p}_u| + |\textbf{c}_u \char`\\ \textbf{p}_u| + |\mathcal{I} \char`\\ \textbf{c}_u|))=O(m)$, which will become clear when converted into a dual form~\cite{li2014top}. 
\textcolor{black}{
We refer the readers to Section 3 of~\cite{li2014top} for the detailed proof regarding the entire process of converting a bipartite ranking into a dual formulation, which in turn gives us the time complexity linear in the number of items.
}

%\vspace{-1ex}
\section{Experiments}
The experiments are designed to answer the following research questions (RQs):
%\vspace{0.5ex}
\begin{enumerate}[leftmargin=0.4in]
	\item[\textbf{RQ1.}] Are click records useful for purchase prediction?
	%	\vspace{1ex}
	\item[\textbf{RQ2.}] Does~\proposetwo~indeed focus on the accuracy near the top? 
	%	\vspace{1ex}
	\item[\textbf{RQ3.}] Is~\proposetwo~robust to unreliable click records?
	%	\vspace{1ex}
	\item[\textbf{RQ4.}] How does the latent dimensionality affect the performance?	
	\item[\textbf{RQ5.}] Does~\proposetwo~outperform baselines without compromising the novelty of item recommendations?	
\end{enumerate}
\label{sec:5}
%\vspace{-0.2ex}
\subsection{Experimental Settings}
\label{sec:5.1}
\noindent\textbf{Dataset. }
We evaluated our proposed method on two real-world datasets each of which contains both purchase records and click records for the same set of users. 
The RecSys2015 dataset\footnote{http://2015.recsyschallenge.com/challenge.html} consists of sessions of click and purchase sequences extracted from an e-commerce website, where we regard each session
is as a user. To the best of our knowledge, the RecSys2015 dataset is the only public dataset in which a user is provided with both the purchase and click records, and hence we ran experiments on a proprietary dataset from NAVER shopping, which is a web portal that provides a platform for online shopping. We collected users' click and purchase records for three months (Jan. 2017 through Mar. 2017). 
For both datasets, we removed users having fewer than five purchases and twenty clicks. 
Moreover, to filter out possible abusing users and items in both datasets, we removed the top 0.001\% of users and items in terms of the number of observations. After preprocessing, the RecSys2015 dataset contained 30,867 purchase records on 5,869 items and 102,939 click records on 11,071 items from 7,076 users, and the NAVER shopping dataset contained 23,373 purchase records on 6,743 items and 243,908 click records on 10,738 items from 5,317 users.

%\medskip
\smallskip
\noindent\textbf{Methods Compared.}
%\vspace{-0.5ex}
\begin{itemize}[leftmargin=.1in]
	\item \textbf{BPR}\footnote{As a naive approach to incorporating both purchase and click records into BPR given the $(u,i\in\textbf{p}_u,j\in\mathcal{I} \char`\\ \textbf{p}_u)$ triple, we modified the value of $(\hat x_{ui} - \hat x_{uj})$ to $0.5*(\hat x_{ui} - \hat x_{uj})$ for $j\in\textbf{c}_u \char`\\ \textbf{p}_u$ under the assumption that the difference should not be as large as when item $j$ is not clicked at all. However, the performance improvement was not statistically significant, hence we excluded the results for brevity.}~\cite{rendle2009bpr}: A pairwise learning-to-rank method based on the \textit{AMAN} assumption as in Eqn.~\ref{AMAN}. 
	\textcolor{black}{
	\item \textbf{SLIM}~\cite{ning2011slim}: An extension of itemKNN~\cite{sarwar2001item} that models the user’s preference for item $i$ as a weighted combination of the user’s preference for item $j$ and the item similarity between $i$ and $j$. It learns a item-item similarity matrix from the data.
	}
	\item \textbf{CLiMF}~\cite{shi2012climf}: A collaborative ranking method for implicit feedback that directly maximizes Mean Reciprocal Rank (MRR).
	\item \textbf{PMF}~\cite{mnih2008probabilistic}: An MF--based pointwise method that minimizes the rating prediction error. As PMF is a common baseline method for rating prediction, we modify it to model click and purchase records; we assign 1 to clicked items, and 2 to purchased items.
	%	Since we are only given implicit feedback, i.e., click and purchase records, we assign 1 to each clicked item and 2 to each purchased item.
	\item \textbf{eALS}~\cite{he2016fast}: The state-of-the-art sampling--based MF method that samples non-purchased items based on their popularity, which is shown to surpass the uniform weighting scheme~\cite{hu2008collaborative}.
	\item \textbf{GRU4REC}~\cite{hidasi2015session}: State-of-the-art session--based click prediction method based on GRU. As its goal (click prediction) differs from ours (purchase prediction), we added a fully connected layer at the end of last hidden state of GRU4REC for predicting the purchased items.
	\item \textbf{\proposed,~\proposedd,~\proposeddd,~\proposetwo,~\proposethree}: Our proposed methods based on Eqn.~\ref{assumption1},~\ref{assumption2},~\ref{assumption3},~\ref{loss}, and~\ref{height}, respectively. Note that~\proposedd~degenerates to BPR when click records are not provided.
	\item \textbf{\infpushcr}~\cite{christakopoulou2015collaborative}: A collaborative ranking method based on explicit feedback that focuses on the ranking performance at the top. Because this method was originally designed for explicit feedback, we cannot directly compare it with our proposed method. Instead, we treat purchased items as relevant and all the non-purchased items as non-relevant. 
	\item \textbf{\proposemix}: A method that jointly minimizes the objective functions of~\proposetwo \\and~\proposethree~as $(1-\varepsilon)\cdot\mathcal{L}_{\proposetwo}(u) + \varepsilon\cdot\mathcal{L}_{\proposethree}(u)$, where $\varepsilon=0.5$.
\end{itemize}
Since our goal is not the click prediction but the purchase prediction, our baseline competitors are built using the purchase records; except for PMF. In fact, we tried the ``click--based purchase prediction'' for BPR (using only click records instead of purchase records to predict purchases) \textcolor{black}{to see how helpful click records are for purchase prediction. }
However, its performance turned out to be very poor, and hence excluded in the paper. Moreover, we assume that explicit feedback such as ratings are not provided, and thus we compare our methods with implicit feedback--based methods.
%\textcolor{black}{Why not click based purchase prediction?}

\medskip
\noindent\textbf{Evaluation Setting. }
We adopt the \textit{leave-one-out} evaluation, which has been widely used in literature~\cite{he2016vbpr,he2017neural,he2016fast,rendle2009bpr}. More precisely, for both datasets, we 1) chronologically ordered the sequence of purchase data and used the last records as test data and the remainder as training data, and 2) used the click records of the user up to the timestamp of the last purchase in the training data and discarded the rest. 
It is worth mentioning that as our target is purchase prediction, where the goal is to predict an item to be purchased in the future, randomly splitting the dataset is unrealistic. Precisely, if we randomly sample a purchased item for each user (without considering the purchased order) and use the rest of the purchased items for training, we will be predicting a past event by using future events. Therefore, for each user, we held out the latest purchased item as the test data, and thus we cannot apply conventional cross-validation. Instead, we ran the experiments five times with different random seeds for initialization for reliability of the results.

\textcolor{black}{
Predicting users' future purchase among
clicked items is rather a trivial task, and obviously the performance
is expected to be significantly improved by incorporating users’
click record as in our method, because items are purchased from clicked items. Indeed, our method greatly outperformed the competitors under such setting. Therefore, to make the problem more challenging and practical~\cite{rendle2010factorizing}, we evaluate our method on how well it predicts users’
future purchase among non-clicked items. That is to say, the candidate items for recommendation for each user are the items that are neither clicked nor purchased by the user in the past. 
}
%Note that we aim to recommend to users, items that are unknown and new to the users, which makes our problem more challenging and practical~\cite{rendle2010factorizing}. That is to say, the candidate items for recommendation for each user are the items that are neither clicked nor purchased by the user in the past. 
%we evaluated our methods on how well they recommend items that are likely to be purchased by a user among items neither clicked nor purchased by the user in the past. 
%Note that recommending items to users from previously clicked items is a rather trivial task, and obviously the performance is expected to be considerably improved by incorporating users' click records as in our method. Therefore, to make the problem more challenging and practical, we evaluated our methods on how well they recommend items that are likely to be purchased by a user from items neither clicked nor purchased by the user in the past. 

\medskip
\noindent\textbf{Evaluation Metrics. } 
As our objective is to optimize the accuracy at the top, we measured the ranking performance using three metrics that emphasize the accuracy at the top (Recall@$N$, NDCG@$N$, MRR@$N$~\cite{donkers2017sequential,ren2018repeatnet,zhang2018towards}) when $N$ is small, and one that does not (AUC). 
\textcolor{black}{
Moreover, since there is only one relelevant item for each user, and that we are dealing with implicit feedback datasets, MRR and NDCG provide the same insight. However, we included both of them because they are the two most popularly used metrics in recommender system research.
}
Precision is not employed because each user has only one test data in which case Precision is proportional to Recall. 
We do not consider metrics, such as RMSE and MAE, as they are suitable for explicit feedback datasets~\cite{forsati2015pushtrust,koren2009matrix}, but not for implicit feedback datasets.
Recall that we split our data into training/validation/test sets by selecting for each user $u$ a random item to be used for validation $\mathcal{V}_u$ and another for testing $\mathcal{T}_u$. All remaining data is used for training. The predicted ranking is evaluated on $\mathcal{T}_u$ with various ranking metrics.
Metrics used for evaluation are described as follows:
\begin{itemize}
	\item Recall@N: The average of the ratio of all relevant items included in top-N of the recommended list of items for each user.\\
	\begin{equation*}
	Recall@N = \frac{1}{n}\sum\limits_{u \in U} {\frac{{|rel(u,N)|}}{|rel(u)|}}
	\end{equation*}	
	where $rel(u,N)$ denotes the relevant (purchased) items among top-N recommended items, and $rel(u)$ denotes relevant (purchased) items of user $u$ in the test set.
	\item Normalized Discounted Cumulative Gain (NDCG)\\
	\begin{equation*}
	NDCG = \frac{1}{n}\sum\limits_{u \in U} {\frac{{DC{G_u}}}{{IDC{G_u}}}}
	\end{equation*}
	where DCG and IDCG (Ideal DCG) are represented as:
	\begin{equation*}
	DC{G_u} = \sum\limits_{i \in rel(u)} {\frac{1}{{{{\log }_2}(rank_i^u + 1)}}}
	\end{equation*}
	\begin{equation*}
	IDCG_u = \sum\limits_{i \in rel(u)} {\frac{1}{{{{\log }_2}(i + 1)}}}
	\end{equation*}
	where $rank_i^u$ denotes the rank of item $i$ in user $u$'s recommendation list.
	\item Mean Reciprocal Rank (MRR)\\
	\begin{equation*}
	MRR = \frac{1}{n}\sum\limits_{u \in U} {\frac{1}{{ran{k^u}}}}
	\end{equation*}
	where $rank_u$ denotes the first rank of the relevant item in the recommended list of user $u$.
	\item Area Under ROC Curve (AUC) \\
	\textcolor{black}{
	\begin{equation*}
	AUC = \frac{1}{n}\sum\limits_{u \in U} {\frac{1}{{|E(u)|}}\sum\limits_{(i,j) \in {E_u}} {\mathbb{I} [{\hat{x}_{ui}} > {\hat{x}_{uj}}]} } 
	\end{equation*}
}
\textcolor{black}{
\noindent where $E(u)=\left\{(i, j) |(u, i) \in \mathcal{T}_{u} \wedge(u, j) \notin\left(\mathcal{P}_{u} \cup \mathcal{V}_{u} \cup \mathcal{T}_{u}\right)\right\}$ and $\mathbb{I}[\cdot]$ is an indicator function that is equal to 1 if the argument is true. $({\hat{x}_{ui}} > {\hat{x}_{uj}})$ indicates that the rank of item $i$ is higher than that of item $j$ for user $u$.
}
\end{itemize}
\textcolor{black}{
Note that if we were not to focus on the top ranks, and if we had more than one relevant item for each user, it would be more rational to test on deeper cut-offs due to the robustness~\cite{valcarce2018robustness}. But in this work, we mainly focus on the accuracy on the top ranks, and thus test on relatively shallow cut-offs. i.e., $N=10,20$.
}

\textcolor{black}{
In addition to the conventional ranking metrics described above, we also evaluate on the novelty of the item recommendations provided by our method. To this end, we adopt self-information (SI)~\cite{castells2011novelty,zhou2010solving}, which measures the unexpectedness of an item recommendation relative to its global popularity:
\begin{equation}
\text{SI}=\frac{1}{|\mathcal{U}|}\sum\limits_{u \in U}\frac{1}{|L(u)|}\sum\limits_{i \in L(u)}-\log_2\frac{C(i)}{|\mathcal{U}|}
\label{eqn:si}
\end{equation}
where $L(u)$ denotes a list of item recommendations for user $u$, $C(i)$ denotes the number of users that purchased item $i$.
}

\begin{table}[t]
	\centering
	%	\small
	\caption{Best performing hyperparameter values. Note that SLIM has two hyparparameters for regularizations (L1/L2).}
	\label{tab:hyper}
	\begin{tabular}{l|ccc|ccc}
		\specialrule{.1em}{.05em}{.05em}
		Data & \multicolumn{3}{c|}{RecSys2015} & \multicolumn{3}{c}{Naver Shopping} \\
		\specialrule{.1em}{.05em}{.05em}
		Method  & $K$     & $\eta$   & $\lambda$  & $K$      & $\eta$   & $\lambda$   \\
		\specialrule{.1em}{.05em}{.05em}
		eALS    & 160   & 0.01   & 0.1      & 150    & 0.01   & 0.01      \\
		BPR     & 190   & 0.1    & 0.01     & 70     & 0.01   & 0.1       \\
		SLIM    & 150 & - & 0.01/5 & 150 & - & 0.01/3 \\
		CLiMF   & 60    & 0.05   & 0.5      & 190    & 0.1    & 0.1       \\
		PMF     & 160   & 0.01   & 0.1      & 170    & 0.01   & 0.1       \\
		GRU4REC     & 150   & 0.01   & -      & 150    & 0.01   & -       \\
		\infpushcr     & 190   & 0.01   & 0.01      & 180    & 0.1   & 0.01       \\
		\proposedd  & 180   & 0.05   & 0.01     & 200    & 0.01   & 0.01      \\
		\proposetwo  & 180   & 0.01   & 0.01     & 160    & 0.1    & 0.01      \\
		\specialrule{.1em}{.05em}{.05em}
	\end{tabular}
	%	\vspace{-3ex}
\end{table}

\begin{landscape}
\begin{table*}[t]
	\centering
	\small
	%	\begin{minipage}[t]{0.7\textwidth}
	\captionof{table}{Performance comparisons for different methods. (\textit{Imp.}: improvement over the best competitor.)}
	\def\arraystretch{0.8}
	\begin{tabular}{c||CCCCCc||Cc|K}
		\specialrule{.1em}{.05em}{.05em}
		\multicolumn{10}{c}{RecSys2015} \\
		\specialrule{.1em}{.05em}{.05em}
		Metric &  eALS & BPR & SLIM & CLiMF &PMF  & GRU4REC & \proposedd  & \proposetwo  & \textit{Imp.}  \\[-0.1em]
		\specialrule{.1em}{.1em}{.1em} 
		%			{\multirow{10}{*}{\rotatebox[origin=c]{90}{RecSys}}}            
		Recall@10   & {  0.2132\scriptsize$\pm0.0006$} & { 0.2801\scriptsize$\pm0.0011$} & { 0.2417\scriptsize$\pm0.0010$} &{ 0.0015\scriptsize$\pm0.0007$} & { 0.2280\scriptsize$\pm0.0024$} & { 0.1916\scriptsize$\pm0.0018$} & { 0.2955\scriptsize$\pm0.0021$} & { \textbf{0.3119}\scriptsize$\pm0.0047$} & { 11.35\%}  \\[-0.1em]
		Recall@20   & {  0.2549\scriptsize$\pm0.0008$} & { 0.3489\scriptsize$\pm0.0012$} & { 0.2919\scriptsize$\pm0.0007$} &{ 0.0022\scriptsize$\pm0.0008$} & { 0.2883\scriptsize$\pm0.0034$}& { 0.2580\scriptsize$\pm0.0015$} & { 0.3853\scriptsize$\pm0.0011$} & { \textbf{0.3959}\scriptsize$\pm0.0068$} & { 13.47\%}  \\[-0.1em]
		NDCG@10   & {  0.1405\scriptsize$\pm0.0004$} & { 0.1792\scriptsize$\pm0.0010$} & { 0.1622\scriptsize$\pm0.0011$} &{ 0.0007\scriptsize$\pm0.0003$} & { 0.1418\scriptsize$\pm0.0019$} & { 0.0895\scriptsize$\pm0.0010$} &  { 0.1778\scriptsize$\pm0.0017$} & { \textbf{0.1913}\scriptsize$\pm0.0050$} & { 6.75\%}  \\[-0.1em]
		NDCG@20   & {  0.1499\scriptsize$\pm0.0006$} & { 0.1966\scriptsize$\pm0.0009$} & { 0.1750\scriptsize$\pm0.0014$} &{ 0.0008\scriptsize$\pm0.0003$} & { 0.1566\scriptsize$\pm0.0017$} & { 0.1063\scriptsize$\pm0.0007$} & { 0.2006\scriptsize$\pm0.0011$} & { \textbf{0.2125}\scriptsize$\pm0.0055$} & { 8.09\%}  \\[-0.1em]
		MRR@10   & {  0.1178\scriptsize$\pm0.0005$} & { 0.1483\scriptsize$\pm0.0009$} & { 0.1374\scriptsize$\pm0.0013$} &{ 0.0005\scriptsize$\pm0.0003$} & { 0.1154\scriptsize$\pm0.0020$} & { 0.0585\scriptsize$\pm0.0009$} & { 0.1425\scriptsize$\pm0.0012$} & { \textbf{0.1542}\scriptsize$\pm0.0051$} & { 3.98\%}  \\[-0.1em]
		MRR@20   & {  0.1204\scriptsize$\pm0.0004$} & { 0.1531\scriptsize$\pm0.0009$} & { 0.1409\scriptsize$\pm0.0002$} &{ 0.0006\scriptsize$\pm0.0003$} & { 0.1194\scriptsize$\pm0.0019$} & { 0.0631\scriptsize$\pm0.0008$} & { 0.1483\scriptsize$\pm0.0014$} & { \textbf{0.1600}\scriptsize$\pm0.0052$} & { 4.51\%}  \\[-0.1em]
		AUC    & {  0.7914\scriptsize$\pm0.0006$} & { 0.8714\scriptsize$\pm0.0005$} & { 0.7012\scriptsize$\pm0.0213$} &{ 0.5523\scriptsize$\pm0.0380$} & { 0.8688\scriptsize$\pm0.0130$} & { 0.7957\scriptsize$\pm0.0009$} &{ \textbf{0.9083}\scriptsize$\pm0.0009$} & { 0.8771\scriptsize$\pm0.0008$} & { 4.23\%}  \\[-0.1em]
		\specialrule{.1em}{.1em}{.1em}
		\multicolumn{10}{c}{NAVER Shopping} \\
		\specialrule{.1em}{.05em}{.05em}
		Recall@10   & { 0.0475\scriptsize$\pm0.0003$} & { 0.0608\scriptsize$\pm0.0010$} & { 0.02914\scriptsize$\pm0.0031$} &{ 0.0076\scriptsize$\pm0.0038$} & { 0.0360\scriptsize$\pm0.0017$} & { 0.0466\scriptsize$\pm0.0022$} & { 0.0618\scriptsize$\pm0.0022$} & { \textbf{0.0690}\scriptsize$\pm0.0015$} & { 13.49\%}  \\[-0.1em]
		Recall@20   & { 0.0739\scriptsize$\pm0.0005$} & { 0.1031\scriptsize$\pm0.0011$} & { 0.0473\scriptsize$\pm0.0012$} &{ 0.0115\scriptsize$\pm0.0039$} & { 0.0646\scriptsize$\pm0.0030$} & { 0.0750\scriptsize$\pm0.0036$} &{ 0.1074\scriptsize$\pm0.0024$} & { \textbf{0.1130}\scriptsize$\pm0.0029$} & { 9.60\%}  \\[-0.1em]
		NDCG@10   & { 0.0222\scriptsize$\pm0.0001$} & { 0.0311\scriptsize$\pm0.0004$} &  {0.0142\scriptsize$\pm0.0014$} &{ 0.0040\scriptsize$\pm0.0029$} & { 0.0180\scriptsize$\pm0.0005$} & { 0.0220\scriptsize$\pm0.0015$} &{ 0.0331\scriptsize$\pm0.0004$} & { \textbf{0.0350}\scriptsize$\pm0.0007$} & { 12.54\%}  \\[-0.1em]
		NDCG@20   & { 0.0288\scriptsize$\pm0.0002$} & { 0.0417\scriptsize$\pm0.0007$} & { 0.0187\scriptsize$\pm0.0007$} &{ 0.0049\scriptsize$\pm0.0025$} & { 0.0244\scriptsize$\pm0.0009$} &{ 0.0291\scriptsize$\pm0.0017$} &{ 0.0443\scriptsize$\pm0.0003$} & { \textbf{0.0460}\scriptsize$\pm0.0008$} & { 10.31\%}  \\[-0.1em]
		MRR@10   & { 0.0147\scriptsize$\pm0.0001$} & { 0.0224\scriptsize$\pm0.0007$} & { 0.0097\scriptsize$\pm0.0008$} &{ 0.0029\scriptsize$\pm0.0029$} & { 0.0126\scriptsize$\pm0.0005$} &{ 0.0147\scriptsize$\pm0.0013$} &{ 0.0233\scriptsize$\pm0.0003$} & { \textbf{0.0244}\scriptsize$\pm0.0010$} & { 8.93\%}  \\[-0.1em]
		MRR@20   & { 0.0165\scriptsize$\pm0.0001$} & { 0.0252\scriptsize$\pm0.0008$} & { 0.0109\scriptsize$\pm0.0006$} &{ 0.0032\scriptsize$\pm0.0028$} & { 0.0143\scriptsize$\pm0.0006$} &{ 0.0166\scriptsize$\pm0.0013$} &{ 0.0263\scriptsize$\pm0.0004$} & { \textbf{0.0274}\scriptsize$\pm0.0010$} & { 8.73\%}  \\[-0.1em]
		AUC    & { 0.7371\scriptsize$\pm0.0156$} & { 0.8622\scriptsize$\pm0.0018$} & { 0.6340\scriptsize$\pm0.0212$} &{ 0.7942\scriptsize$\pm0.0156$} & { 0.8520\scriptsize$\pm0.0003$} & { 0.7233\scriptsize$\pm0.0011$} &{ \textbf{0.9420}\scriptsize$\pm0.0005$} & { 0.8515\scriptsize$\pm0.0014$} & { 9.26\%}  \\[-0.1em]
		\specialrule{.1em}{.05em}{.05em}
	\end{tabular}
	%	\vspace{-4ex}
	\label{tab:overall}
\end{table*}
\end{landscape}

\medskip
\noindent\textbf{Implementation. }
To make fair comparisons, we built on one of the most widely used library for recommender systems, called LibRec\footnote{https://www.librec.net/}. We used their implementations of BPR, \textcolor{black}{SLIM}, CLiMF and PMF, and implemented eALS, variants of~\propose~and~\proposetwo. We used PyTorch~\cite{paszke2017automatic} to implement GRU4REC.

\medskip
\noindent\textbf{Parameters. }
For all baselines, we tuned the hyperparameters by performing grid searches with  $K\in\{10,20,...,200\}$, and $\eta$ (learning rate)$, \lambda\in\{0.01,0.05,0.1\}$. For each user, we used the last purchased item as test data and the remainder as training data. Therefore, conventional cross-validation is not applicable here, as the data must be split based on time. Instead, we made a validation dataset by using the last purchased item of training data, and performed grid search on the validation dataset for five times with different random seeds for initialization to find the best hyperparameters. The best performing hyperparameter values for each method found by grid search on the validation dataset are summarized in Table~\ref{tab:hyper}. 
For experiments, we report the mean and the standard deviation over five runs with different random seeds for initialization; the standard deviations in graphs are displayed as error bars. 
%The reported results in our experiments are averaged over five runs with different random seeds for initialization. The standard deviations in graphs are displayed as error bars. 

\begin{table}[h]
	%	\vspace{-1ex}
	\centering
	\captionof{table}{Comparisons of different assumptions for~\propose.}
	%	\vspace{-2ex}
	\label{tab:P3S}
	%	\begin{tabular}{c|c||ccc}
	%		\specialrule{.1em}{.05em}{.05em}
	%		Data                                    & Metric &  \proposed  & \proposedd  & \proposeddd  \\
	%		\specialrule{.1em}{.1em}{.1em}
	%		{\multirow{4}{*}{\rotatebox[origin=c]{90}{RecSys}}}
	%%		& P@10   & {0.0277${(0.0001)}$} & {\textbf{0.0296}${(0.0002)}$} & {0.0002${(0.0000)}$}  \\[-0.15em]
	%		& R@10   & {0.2772${(0.0009)}$} & {\textbf{0.2955}${(0.0021)}$} & {0.0017${(0.0004)}$}  \\[-0.15em]
	%		& N@10   & {0.1773${(0.0005)}$} & {\textbf{0.1778}${(0.0017)}$} & {0.0014${(0.0003)}$}  \\[-0.15em]
	%		& M@10   & {\textbf{0.1463}${(0.0006)}$} & {0.1425${(0.0012)}$} & {0.0011${(0.0003)}$}  \\[-0.12em]
	%		& AUC    & {0.8716${(0.0006)}$} & {\textbf{0.9083}${(0.0009)}$} & {0.4602${(0.0035)}$}  \\[-0.15em]
	%		\specialrule{.1em}{.1em}{.1em}
	%		%			\specialrule{.1em}{.1em}{.1em}
	%		{\multirow{4}{*}{\rotatebox[origin=c]{90}{Naver Shopping}}}		   
	%%		& P@10   & {0.0060${(0.0002)}$} & {\textbf{0.0062}${(0.0002)}$} & {0.0003${(0.0001)}$}  \\[-0.15em]
	%		& R@10   & {0.0602${(0.0016)}$} & {\textbf{0.0618}${(0.0022)}$} & {0.0028${(0.0010)}$}  \\[-0.15em]
	%		& N@10   & {0.0298${(0.0006)}$} & {\textbf{0.0331}${(0.0004)}$} & {0.0013${(0.0007)}$}  \\[-0.15em]
	%		& M@10   & {0.0215${(0.0006)}$} & {\textbf{0.0233}${(0.0003)}$} & {0.0009${(0.0006)}$}  \\[-0.15em]
	%		& AUC    & {0.8765${(0.0029)}$} & {\textbf{0.9420}${(0.0005)}$} & {0.4948${(0.0050)}$}  \\[-0.1em]
	%		\specialrule{.1em}{.05em}{.05em}
	%	\end{tabular}
	%		\def\arraystretch{0.95}
	\begin{tabular}{c|c||ccc}
		\specialrule{.1em}{.05em}{.05em}
		Data                                    & Metric &  \proposed  & \proposedd  & \proposeddd  \\[-0.15em]
		\specialrule{.1em}{.1em}{.1em}
		{\multirow{4}{*}{\rotatebox{90}{\small RecSys2015}}}
		%	& P@10   & {0.0277$\pm0.0001$} & {\textbf{0.0296}$\pm0.0002$} & {0.0002$\pm0.0000$}  \\[-0.15em]
		& Recall@10   & {0.2772$\pm0.0009$} & {\textbf{0.2955}$\pm0.0021$} & {0.0017$\pm0.0004$}  \\[-0.15em]
		& NDCG@10   & {0.1773$\pm0.0005$} & {\textbf{0.1778}$\pm0.0017$} & {0.0014$\pm0.0003$}  \\[-0.15em]
		& MRR@10   & {\textbf{0.1463}$\pm0.0006$} & {0.1425$\pm0.0012$} & {0.0011$\pm0.0003$}  \\[-0.12em]
		& AUC    & {0.8716$\pm0.0006$} & {\textbf{0.9083}$\pm0.0009$} & {0.4602$\pm0.0035$}  \\[-0.15em]
		\specialrule{.1em}{.1em}{.1em}
		%			\specialrule{.1em}{.1em}{.1em}
		{\multirow{4}{*}{\rotatebox{90}{\small Naver Shopping}}}		   
		%	& P@10   & {0.0060$\pm0.0002$} & {\textbf{0.0062}$\pm0.0002$} & {0.0003$\pm0.0001$}  \\[-0.15em]
		& Recall@10   & {0.0602$\pm0.0016$} & {\textbf{0.0618}$\pm0.0022$} & {0.0028$\pm0.0010$}  \\[-0.15em]
		& NDCG@10   & {0.0298$\pm0.0006$} & {\textbf{0.0331}$\pm0.0004$} & {0.0013$\pm0.0007$}  \\[-0.15em]
		& MRR@10   & {0.0215$\pm0.0006$} & {\textbf{0.0233}$\pm0.0003$} & {0.0009$\pm0.0006$}  \\[-0.15em]
		& AUC    & {0.8765$\pm0.0029$} & {\textbf{0.9420}$\pm0.0005$} & {0.4948$\pm0.0050$}  \\[-0.1em]
		\specialrule{.1em}{.05em}{.05em}
	\end{tabular}
	%	\vspace{-3ex}
\end{table}

\subsection{Performance Analysis}
\noindent\textbf{RQ1) Usefulness of Click Records. }
We begin by showing which of our model assumptions (Eqn.~\ref{assumption1},\ref{assumption2}, or~\ref{assumption3}) performed the best in Table~\ref{tab:P3S}. We observe that~\proposedd, which is based on \scshape \textbf{Assumption 2}\normalfont~(Eqn.~\ref{assumption2}), generally outperforms~\proposed~and~\proposeddd~on both datasets, with~\proposeddd~performing extremely poorly. This implies that \textit{clicked-but-not-purchased} items of a user are definitely more helpful 
in eliciting the user's preference than \textit{non-clicked} items, and thus this relationship should be taken into account for purchase prediction.
%	Moreover, we speculate that the poor performance of~\proposeddd~results from the diverse composition of item categories of non-clicked items.
%	In other words, non-clicked items are comprised of more diverse item categories compared with clicked-but-not-purchased items. 
%	For a statement in \scshape \textbf{Assumption 3}\normalfont, which is ``\textit{non-clicked} items are more preferred by a user to \textit{clicked-but-not-purchased} items because the user discarded the \textit{clicked-but-not-purchased} items whereas a large portion of \textit{non-clicked} items are not known to the user yet'' to be valid, the category of non-clicked items should not be far from that of clicked-but-not-purchased items, otherwise such ordering of preferences among itemsets is meaningless.

%\begin{figure}[t]
%	\captionsetup[subfigure]{aboveskip=3pt}
%	\centering
%	\begin{subfigure}[b]{0.47\textwidth}        %% or \columnwidth
%		\centering
%		\includegraphics[width=\linewidth]{./figure/recsys_topN_top.eps}
%		\caption{Performance at the top ($N=1\sim10$)}
%		\label{fig:topN-A}
%%		\vspace{-0.5ex}
%	\end{subfigure}
%	\begin{subfigure}[b]{0.47\textwidth}        %% or \columnwidth
%		\centering
%		\includegraphics[width=\linewidth]{./figure/recsys_topN_bottom.eps}
%		\caption{Performance at the top ($N=15\sim500$)}
%		\label{fig:topN-B}
%	\end{subfigure}
%%	\vspace{-2ex}
%	\caption{Performance comparisons on RecSys2015 dataset.}
%	\label{fig:topN}
%%	\vspace{-4ex}
%\end{figure}

Given that~\proposedd~is the right choice among the~\propose s, we now compare its performance with state-of-the-art implicit feedback based--recommendation methods in Table~\ref{tab:overall}. We observe that \proposedd~
consistently outperformed the purchase record--based baselines, i.e., eALS, CLiMF and BPR, with very few exceptions. This indicates that when click records are properly combined with purchase records to define the order of user preferences among non-purchased items, we can complement the missing user--item interactions of purchase records, which eventually leads to better recommendation quality.
Other observations from Table~\ref{tab:overall} are as follows: \textbf{1)} BPR consistently outperformed PMF. Although PMF utilizes both the click and purchase records for making recommendations, it performed worse than BPR, which leverages only the purchase records. This indicates that we should meticulously design a method when jointly modeling both the click and purchase records, which in fact was one of the objectives for this study.
\textbf{2)} GRU4REC, which is a click-based purchase prediction method, generally performs worse than other methods based on either \textit{only purchase} or \textit{both click and purchase}. This shows that while click records can be directly used for click prediction as in~\cite{hidasi2015session}, using only click records for purchase prediction limits the prediction performance because click records are relatively weaker signal than purchases, which corroborates the benefit of our model assumptions for purchase prediction. However, we note that GRU4REC performs better than click--based version of BPR (not shown in Table~\ref{tab:overall}), confirming the advantage of considering the sequential information of click records by using GRU.
\textbf{3)} BPR consistently outperformed eALS. This demonstrates the superiority of the pairwise learning-to-rank method upon which our method is built. 
\textbf{4)} Although we expected CLiMF to perform better than BPR as in the original study~\cite{shi2012climf}, its performance turned out to be very poor. We attribute this poor performance to the difference in the target task, which leads to a different model formulation. 
More precisely, CLiMF ignores all the missing user--item interactions and only considers positive feedback, and we argue that this can be effective for tasks with sufficient positive feedback such as friend recommendations (about 70 friends per user on average for the datasets used by~\cite{shi2012climf}).
%	However, for the task of user purchase prediction where the number of positive user feedback is extremely small (4 purchased items per user in average for datasets used in this work), such formulation is not appropriate, which is explained by the poor performance of CLiMF in our task. 
However, the poor performance of CLiMF in our task implies that this formulation is not appropriate for purchase prediction where the amount of positive feedback is usually small (four purchased items per user on average for the datasets used in this work). 
In addition, the poor performance of CLiMF compared with BPR, which leverages the negative feedback, corroborates the importance of leveraging negative feedback in different target tasks.
%	Such formulation is feasible when a large number of positive feedback is given as in the social network datasets used to evaluate CLiMF (70 friends per user in average). However, in case of e-commerce domain, the number of positive feedback (4 purchased items per user in average for datasets used in this work) is usually considerably less than the number of friends in social network. Due to the above reasons, we conjecture that the performance of CLiMF is inevitably degraded in our task.
%	CLiMF shows extremely poor performance. We speculate that such poor performance of CLiMF on our task results from the difference in the target task that leads to different evaluation protocol. Precisely, CLiMF is originally designed for the task of friend recommendation in social networks where the number of friends (about 70 friends per user in average for datasets used in~\cite{shi2012climf}), which are analogous to items in e-commerce, usually considerably outnumbers the number purchased items (about 4 purchased items per user in average for datasets used in this work) in e-commerce stores. Note that the evaluation protocol used in the paper that proposed CLiMF~\cite{shi2012climf} was to use a fixed number of instances for each user as training data and the rest as the test data, and thus many more instances for test compared with our protocol where there is only one instance in the test data for each user. Such difference leads to relatively poor performance of CLiMF when applied to our task.
\textcolor{black}
{5) SLIM performs relatively better on RecSys2015 dataset compared with NAVER dataset. Precisely, on Recsys2015 dataset, SLIM performs the second best among the baselines, whereas it performs worse on NAVER dataset. This is mainly because SLIM can only model relations between items that have been co-purchased by at least some users, which implies that SLIM performs well on dense datasets than on sparse datasets~\cite{kabbur2013fism}. In the same vein, we attribute the inferior performance of SLIM compared with BPR to the sparseness of our datasets.
}

\begin{figure}[h]
	\centering
	\includegraphics[width=0.5\textwidth]{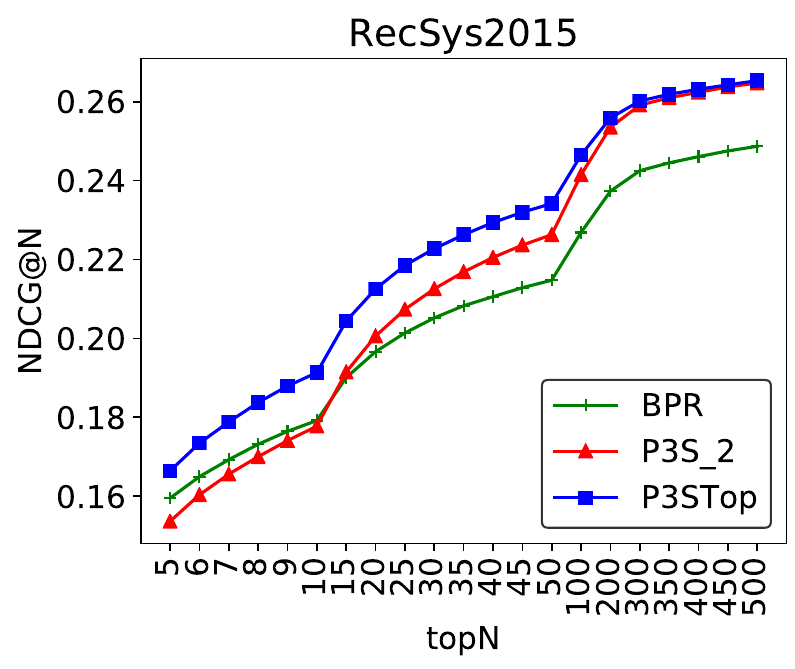}
	\caption{Comparisons over various $N$s (RecSys2015).}
	\label{fig:topN}
\end{figure}

\medskip
\noindent\textbf{RQ2) Focus on the Accuracy of the Top-Ranked Items. }
We observe from Figure~\ref{fig:topN} that although \proposedd~outperforms BPR for relatively large $N$s ($N\geq20$), their performance is similar to BPR for smaller $N$s less than 10; in this range, BPR even performs better than~\proposedd.
%their performance gap becomes even larger as we go down the recommendation list as shown in Figure~\ref{fig:topN}. For example, in case of Recall, while BPR performed better from $N=$ 1 to 5 (Figure~\ref{fig:topN-A}), the performance of~\proposedd~exceeded the performance of BPR from $N=$ 6 to 500 (Figure~\ref{fig:topN-A},~\ref{fig:topN-B}). 
However, providing accurate recommendations in the lower part of the recommendation list as~\proposedd~is not desired for e-commerce stores because \textit{users are mainly interested in the top-ranked items} in practice~\cite{top}. 
Recall that the objective of our ultimate proposed method,~\proposetwo, is to focus on the accuracy of the top-ranked items.
%to provide accurate results near the top was one of our motivations, which
%Hence, which was the motivation of our ultimate proposed method~\proposetwo.
%one of the motivations of this work is to provide more accurate recommendation results near the top, by adopting the push/pull mechanism.
%Hence, we now evaluate the performance of~\proposetwo, which is designed to optimize the accuracy of the top-ranked items. 
The following observations are made regarding the performance of~\proposetwo:
\textbf{1)} From Table~\ref{tab:overall}, in terms of Recall, NDCG and MRR,~\proposetwo~considerably outperforms all competitors including~\proposedd~for $N=10,20$, which are relatively small $N$s.
\textbf{2)} More importantly, for even smaller $N$s less than 10 (Figure~\ref{fig:topN})\footnote{\proposetwo~ outperforms BPR from top-3, but excluded for the clarity of the graph.}, we observe that~\proposetwo~still outperforms both BPR and~\proposedd, whereas for large $N$s ($N\geq30$), the performance gap between~\proposedd~ and~\proposetwo~starts to get smaller, and the performance of~\proposetwo~almost equals to that of~\proposedd~above $N=$ 300. This implies that~\proposetwo~focuses on the accuracy of the top-ranked items ($N\leq20$) 
at the expense of the accuracy
%while sacrificing the accuracy 
%of the items 
in the lower part of the recommendation list ($N\geq30$), which answers RQ2. We observed similar results for other metrics as well. 
\textbf{3)} Above results are corroborated by the performance in terms of AUC, a metric that treats a mistake in the higher part of the recommendation list as equal to one the lower part. More precisely,~\proposedd~consistently outperforms~\proposetwo~in both datasets in terms of AUC, which implies that~\proposedd~provides a more balanced recommendation list; this conversely shows that~\proposedd~does not particularly focus on the top. We attribute this performance to the fact that~\proposedd~is built upon the BPR model, whose objective is to optimize for the AUC metric (Eqn.~\ref{likelihood_P3S2}). 
%As the goal of~\proposetwo~is to optimize for the accuracy of the top-ranked items, AUC is not a suitable metric by which~\proposetwo~should be evaluated.

\begin{table}[t]
	\centering
	\caption{Comparisons among ``\textit{push}'' algorithms.}
	%	\vspace{-1ex}
	\label{tab:P3SPush}
	\small
	\begin{tabular}{c||cccc}
		\specialrule{.1em}{.05em}{.05em}
		\multicolumn{5}{c}{RecSys2015} \\
		\specialrule{.1em}{.05em}{.05em}
		Metric & \infpushcr &  \proposetwo  & \proposethree  & \proposemix  \\
		\specialrule{.1em}{.1em}{.1em}  
		%	P@10   & {0.0163$\pm0.0002$} & {\textbf{0.0312}$\pm0.0005$} & {0.0189$\pm0.0003$} & {0.0184$\pm0.0002$}  \\[-0.13em]
		R@10   & {0.1631$\pm0.0018$} & {\textbf{0.3119}$\pm0.0047$} & {0.1891$\pm0.0034$} & {0.1842$\pm0.0024$}  \\[-0.1em]
		N@10   & {0.0818$\pm0.0011$} & {\textbf{0.1913}$\pm0.0050$} & {0.1021$\pm0.0032$} & {0.0989$\pm0.0015$}  \\[-0.1em]
		M@10   & {0.0577$\pm0.0011$} & {\textbf{0.1542}$\pm0.0051$} & {0.0760$\pm0.0032$} & {0.0731$\pm0.0015$}  \\[-0.1em]
%		AUC    & {0.7401$\pm0.0019$} & {\textbf{0.9039}$\pm0.0008$} & {0.8806$\pm0.0016$} & {0.8648$\pm0.0013$}  \\[-0.1em]
		%			\specialrule{.1em}{.1em}{.1em}
		\specialrule{.1em}{.1em}{.1em}
		\multicolumn{5}{c}{Naver Shopping} \\
		\specialrule{.1em}{.05em}{.05em}
		%	P@10   & {0.0017$\pm0.0001$} & {\textbf{0.0069}$\pm0.0002$} & {0.0049$\pm0.0003$} & {0.0045$\pm0.0002$}  \\[-0.1em]
		R@10   & {0.0173$\pm0.0015$} & {\textbf{0.0690}$\pm0.0015$} & {0.0487$\pm0.0026$} & {0.0446$\pm0.0018$}  \\[-0.1em]
		N@10   & {0.0082$\pm0.0008$} & {\textbf{0.0350}$\pm0.0007$} & {0.0212$\pm0.0013$} & {0.0227$\pm0.0014$}  \\[-0.1em]
		M@10   & {0.0053$\pm0.0007$} & {\textbf{0.0244}$\pm0.0010$} & {0.0152$\pm0.0011$} & {0.0164$\pm0.0012$}  \\[-0.1em]
%		AUC    & {0.6010$\pm0.0026$} & {\textbf{0.9313}$\pm0.0014$} & {0.8627$\pm0.0028$} & {0.8287$\pm0.0013$}  \\[-0.1em]
		\specialrule{.1em}{.05em}{.05em}
	\end{tabular}
	%\vspace{-2ex}
\end{table}

\smallskip
\noindent\textbf{RQ3) Robustness to Unreliable Click Records. }
%	RQ3) Comparison between ``\textit{Push}'' Algorithms. 
Table~\ref{tab:P3SPush} shows the comparisons among ``\textit{push}'' algorithms that focus on accuracy of the top-ranked items. 
%The performance of~\proposethree, which is an enhanced version of~\infpushcr, is poor compared with~\proposetwo,
Our proposed method~\proposetwo~ (Figure~\ref{fig:bound}c), which places more emphasis on positive items than on negative items, considerably outperformed~\proposethree \\(Figure~\ref{fig:bound}d), which places more emphasis on negative items than on positive items. This verifies that resorting to negative items deteriorates the recommendation performance implying that click records are indeed relatively more unreliable than purchase records, which answers RQ3.

Other observations from Table~\ref{tab:P3SPush} are as follows;
Note that~\infpushcr~\cite{christakopoulou2015collaborative}~is a collaborative ranking method based on explicit feedback that pulls the incorrectly ranked relevant items above non-relevant items.
%pushes the non-relevant items below relevant items. 
%Because this method was originally designed for explicit feedback, we cannot directly compare it with our proposed method. Instead, we treat purchased items as relevant and all the non-purchased items as non-relevant. 
\textbf{1)} Although~\proposethree~performs worse than~\proposetwo, \proposethree~slightly outperforms~\infpushcr, which verifies again that defining the order of user preferences among non-purchased items as specified in \scshape \textbf{Assumption 2}\normalfont~is indeed beneficial. 
%2) \proposetwo~considerably outperformed the other ``push'' algorithms. This again demonstrates that click records are noisy and that we should rely more on purchase records to make the model robust to noisy click records.
\textbf{2)} The performance of~\infpushcr~is very poor compared with not only other ``\textit{push}'' algorithms but also the competitors listed in Table~\ref{tab:overall}. Recall that~\infpushcr~is distinguished from~\proposethree~in that the underlying assumption of~\proposethree, i.e., \scshape \textbf{Assumption 2}\normalfont, is replaced with the \textit{AMAN} assumption. 
Hence, similar to~\proposethree~explained in Section~\ref{sec:alt}, the lower bound of purchased items set by the top-ranked non-purchased item should be high enough so that pulling up purchased items above it yields desired results. However, the poor performance of~\infpushcr~implies that non-purchased items should not be equally considered as negative, and that we need to define the order of user preferences among non-purchased items by taking into account  \textit{clicked-but-not-purchased} items.
\textbf{3)} The performance of~\proposemix~is worse than both~\proposetwo~and~\proposethree, which implies that jointly learning these two methods provides no benefit owing to the unreliableness of click records.

\begin{figure}[h]
	\captionsetup[subfigure]{aboveskip=0pt}
	\centering
	\begin{subfigure}[b]{0.9\textwidth}        %% or \columnwidth
		\centering
		\includegraphics[width=\linewidth]{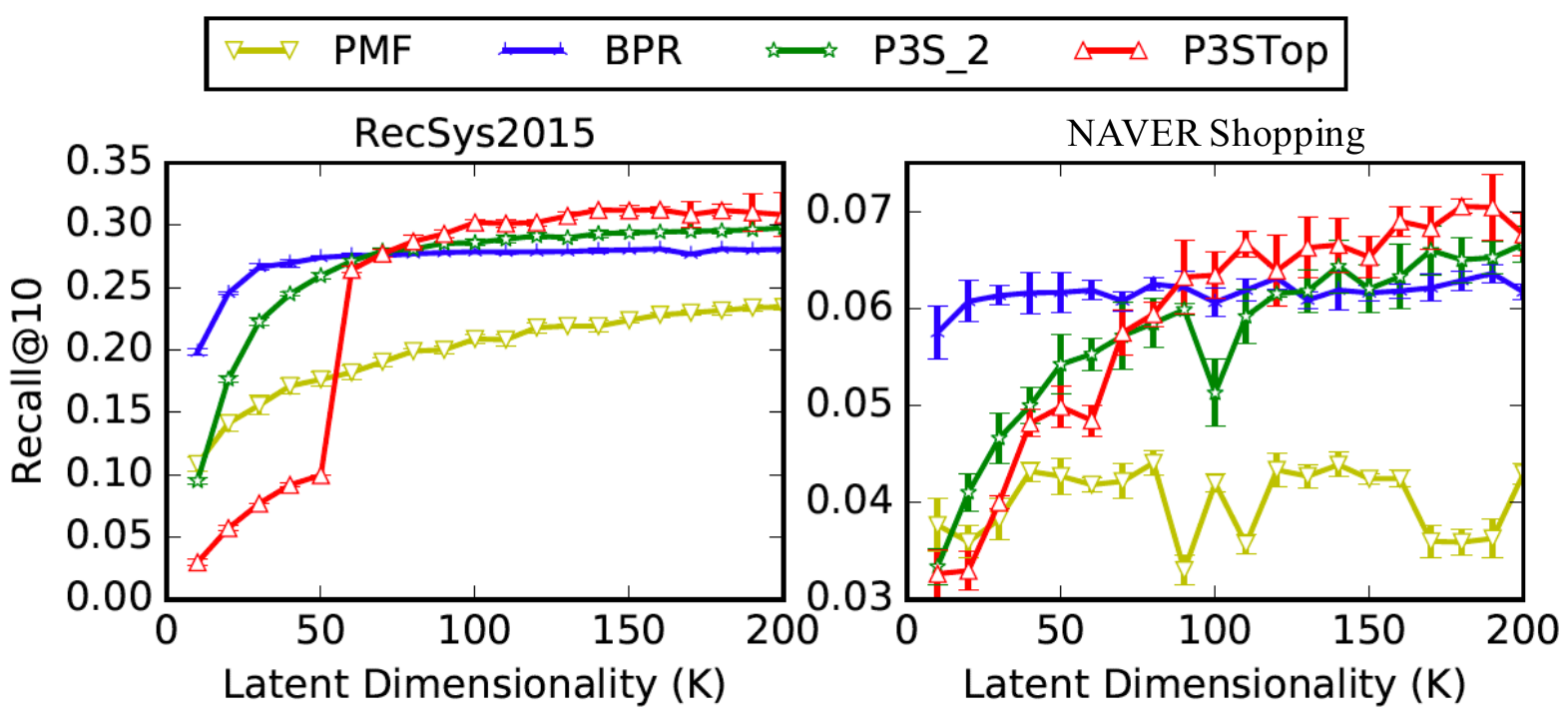}
		%		\vspace{-1.5ex}
		%		\caption{Performance in terms of Recall@10.}
	\end{subfigure}
	%	\begin{subfigure}[b]{0.4\textwidth}        %% or \columnwidth
	%		\centering
	%		\includegraphics[width=\linewidth]{./figure/dim_revision_ndcg.eps}
	%		\caption{Performance in terms of NDCG@10.}
	%	\end{subfigure}
	%\vspace{-2ex}
	\caption{Recall@10 w.r.t. various $K$s on both datasets.}
	\label{fig:dim}
	%	\vspace{-2ex}
\end{figure}

\medskip
\noindent\textbf{RQ4) Dimensionality Analysis. }
Figure~\ref{fig:dim} shows the impact of the number of latent dimensions $K$ on Recall for both datasets. While the performance of every method improved as $K$ increased, the performance improvements were more significant for~\proposedd~and~\proposetwo.	
We attribute this improvement to the fact that these methods need a larger model capacity than the rest of the methods because multiple relationships among itemsets are considered.

\begin{table}[h]
	\centering
	\begin{tabular}{c||cc}
		Self-information & BPR    & \proposetwo  \\
		\hline
		\hline
		RecSys2015       & 8.8930 & 9.3439 \\
		NAVER Shopping   & 8.4382 & 10.0342
	\end{tabular}
\caption{Mean self-information of top-10 recommended items on both datasets.}
\label{tab:si}
\end{table}

\medskip
\textcolor{black}{
\noindent\textbf{RQ5) Preserving the novelty of recommendation.}
While it is important to provide accurate recommendations to users, another aspect of a successful recommender system that should not be neglected is the novelty of recommendations, as accuracy alone does not always result in user satisfaction~\cite{kotkov2018investigating}. In this respect, we compare the self-information (Equation~\ref{eqn:si} of top-10 recommended items of BPR with those of\textcolor{black}{~\proposetwo}~in Table~\ref{tab:si}. We observe that~\proposetwo~not only provides accurate recommendations compared with BPR, but also novel recommendations. We argue that this is mainly due to the fact that click records help reduce the dependence on the item popularity.
}

\section{Related Work}
%\noindent\textbf{Recommender Systems with Implicit Feedback.}
\subsection{Recommender Systems with Implicit Feedback}
Although explicit feedback, such as rating, is a valuable source of information that reveals user preferences, it is difficult to obtain a large quantity of such data. Hence, the vast majority of work has focused on eliciting user preferences from implicit feedback such as bookmarks~\cite{zhao2014leveraging}, item purchases~\cite{rendle2009bpr,he2016vbpr}, and TV channel tuning history~\cite{pan2008one}. These methods adopted the MF technique to model the preference of users on items~\cite{koren2009matrix}.
Specifically, Hu~\etal~proposed WMF that~\cite{hu2008collaborative} introduced the concept of confidence to measure the influence of observed items and unobserved items on users' preferences.
%\begin{equation}
%	\label{eqn:wmf}
%	\mathcal{L}_{WMF} = \sum_{u\in \mathcal{U}}\sum_{i\in \mathcal{I}}c_{ui}(r_{ui}-\alpha_u^T\beta_i)^2+ \lambda_\Theta \norm{\Theta}_F^2
%\end{equation}
%where $r_{ij}=1$ if user $u$ interacted with item $i$, and 0 otherwise. $\Theta$ is the set of parameters to be learned. $\alpha_u\in\mathbb{R}^K$ and $\beta_i\in\mathbb{R}^K$ represent the latent factors of user $u$ and item $i$, respectively, where $K$ is the latent dimensionality. $\mathcal{U}$ and $\mathcal{I}$ are set of users and items, respectively. Here, $c_{ui}$ measures the confidence of $r_{ui}$ being equal to 1, which is determined in advance.
Later, various sampling strategies to generate negative examples from unobserved items were proposed~\cite{pan2008one,he2016fast,rendle2014improving}. 
%Moreover, Rendle~\etal~proposed BPR~\cite{rendle2009bpr}, which is a pairwise learning-to-rank method whose objective is based on pairwise comparisons between observed items and unobserved items under the assumption of Eqn.~\ref{AMAN}.
Moreover, several pairwise learning-to-rank methods~\cite{rendle2009bpr,pan2013gbpr,weston2011wsabie,li2016relaxed} based on pairwise comparisons between observed items and unobserved items have been proposed.
%Rendle~\etal~proposed BPR~\cite{rendle2009bpr}, which is a pairwise learning-to-rank method whose objective is based on pairwise comparisons between observed items and unobserved items under the assumption of Eqn.~\ref{AMAN}.
%More precisely, the objective function of BPR is formulated based on the assumption of 
%\begin{equation}
%	\label{eqn:bpr}
%	\mathcal{L}_{BPR} = \sum_{u\in \mathcal{U}}\sum_{i\in \textbf{p}_u}\sum_{j\in \mathcal{I}\textbackslash \textbf{p}_u} ln\sigma( \hat x_{ui} - \hat x_{uj}) - \lambda_\Theta ||\Theta||_F^2
%\end{equation}
%\begin{equation}
%\label{eqn:bpr}
%\mathcal{J}_{BPR}(\Theta)=\prod\limits_{u\in\mathcal{U}} \prod\limits_{i\in \textbf{p}_u}\prod\limits_{j\in \mathcal{I} \char`\\ \textbf{p}_u} \Pr[i \succ_u j]
%\end{equation}
However, all the aforementioned methods are based on the \textit{AMAN} assumption or predefined heuristic weights, which limits further performance improvement. 

To cope with the aforementioned challenges, users' social network information has been leveraged. For example, a method introduced by Zhao~\etal~assigned higher ranks to the items that a user's friends prefer than to the items that neither he nor his friends prefer~\cite{zhao2014leveraging}. This work was extended by \textcolor{black}{Wang~\etal\cite{wang2016social}}, who introduced a method that categorizes unobserved items into three groups regarding users' strong and weak ties with other users. However, these methods are only applicable when users' social network information is available, which is usually not the case for most e-commerce stores.
Various other methods that incorporate side information for solving the data sparsity issue of implicit feedback have been proposed: review text~\cite{wang2011collaborative}, item image~\cite{he2016vbpr} and temporal information~\cite{rendle2010factorizing}. However, this line of research is not directly related to our proposed method in that ours does not consider any side information.
Moreover, dwell time~\cite{yi2014beyond} can be used to emphasize more reliable clicks, however, we argue that dwell time is a type of ``temporal'' side information that cannot be readily obtained.
Lastly, Parra~\etal~\cite{parra2011implicit} proposed a parametric model to map implicit feedback to explicit feedback under the assumption that there is some correlation between implicit and explicit feedback. However, it requires a minimal amount of explicit feedback, whereas ours entirely resort to implicit feedback.
\textcolor{black}{
Finally, as alternative approach for the same problem, we could think of feature engineering--based methods. Feature engineering--based methods refer to methods that manually generate features regarding the users and items. However, the generation of features is domain-specific~\cite{cheng2016wide}, labor intensive and insufficient to uncover the underlying properties of data~\cite{sun2015modeling}.
}

\medskip
%\noindent\textbf{Modeling User Behavior.} 
\subsection{Modeling User Behavior}
With the advent of e-commerce, much work has been devoted to understanding behavior of online users~\cite{lo2016understanding,cheng2017predicting}, and specifically to predicting purchase behaviors~\cite{li2015predicting,liu2016repeat}.
As the former line of work, Lo~\etal~\cite{lo2016understanding} studied user activity and purchasing behaviors that vary over time, especially focusing on user purchasing intent. Most recently, Cheng~\etal~\cite{cheng2017predicting} extended Lo~\etal's work~\cite{lo2016understanding} by generalizing their analysis on characterizing the relationship between a user's intent and his behavior. 
Our goal is different in that we focus on predicting users' purchases, rather than predicting users' various intents from their online behaviors.
Meanwhile, as the latter line of work, 
given user demographics and implicit feedback including click record and purchase record, 
Liu~\etal~\cite{liu2016repeat} proposed an ensemble method to predict which customers would return to the same merchant within six months period. They formulated the problem as a classification task and trained various classification methods. While similarly using both purchase record and click record, our task is different in that we aim to predict items that users will purchase rather than to predict repeat buyers. Moreover, Li~\etal~\cite{li2015predicting} proposed a MF--based method that predicts the conversion response of users in display advertising, the goal of which inherently differs from our task. 

%It is worth noting that click records have been used for various tasks such as click-through rate (CTR) prediction in online advertising~\cite{mcmahan2013ad,zhu2010novel,zhang2014sequential} and Twitter~\cite{li2015click}, user intent prediction~\cite{lo2016understanding,cheng2017predicting}, repeat-buyer prediction~\cite{liu2016repeat}, conversion response prediction in display advertising~\cite{li2015predicting}, and session--based click prediction~\cite{hidasi2015session}. However, not much effort has been devoted to purchase prediction, especially to leveraging click records to complement the missing user--item interactions of purchase records as done in this paper.
%To the best of our knowledge, our work is the first attempt to leverage click records to complement the missing user--item interactions of purchase records.
%predict users' future purchase by jointly modeling both users' click and purchase record.

\medskip
%\noindent\textbf{Optimizing the Accuracy at the Top.}
\subsection{Optimizing the Accuracy at the Top}
%Learning-to-rank is a supervised machine learning method that directly builds a ranking list from training data~\cite{liu2009learning}, and has been actively researched in many information retrieval problems such as document retrieval and recommender systems. 
Considering that users are mainly interested in the top-ranked items~\cite{top}, optimizing for the accuracy near the top is of great importance in practice. 
Thanks to the success of the above approaches in general ranking tasks~\cite{narasimhan2013structural,kar2015surrogate,rudin2009p}, they have been recently adopted in the field of recommender systems. Weimer~\etal~proposed CoFiRank~\cite{weimer2008cofi}, which directly optimizes Normalized Discounted Cumulative Gain (NDCG) by minimizing its convex upper bound. Later, Shi~\etal~proposed CLiMF~\cite{shi2012climf}, xCLiMF~\cite{shi2013xclimf}, and GAPfm~\cite{shi2013gapfm}, which optimize mean reciprocal rank, expected reciprocal rank and graded average precision, respectively. Among these methods, CLiMF is based on implicit feedback, whereas others are based on explicit feedback, and thus we compared our proposed method with CLiMF in our experiments in Section~\ref{sec:5}.
Furthermore, Christakopoulou and Banerjee proposed PushCR, which applies p-norm push, infinite push and reverse-height push~\cite{rudin2009p} to a collaborative ranking task in which the ranking loss focuses on the accuracy of the top-ranked items for each user~\cite{christakopoulou2015collaborative}. Hu and Li recently proposed DCR, which focuses on the accuracy at the top by modeling user ratings based on an ordinal classification framework~\cite{hu2017decoupled}.
Lastly, Forsati~\etal~\cite{forsati2015pushtrust} and Rafailidis and Crestani~\cite{rafailidis2016joint} incorporated user social network data as side information to enhance the accuracy at the top. However, these methods cannot be directly compared with ours because 1) PushCR and DCR consider explicit feedback, whereas ours is solely based on implicit feedback, and 2) the latter works incorporate side information related to users, whereas ours does not.

\medskip
%\noindent\textbf{Position Bias of Click Models.}
\subsection{Position Bias of Click Models}
Position bias is a fundamental problem pertaining to click records, where users tend to click on higher ranked items regardless of their relevance~\cite{richardson2007predicting,chen2012position}. To tackle the position bias issue for CTR prediction, previous click models assume that the click probability depends on the probability of examining a position, and the relevance of the document displayed at that position. However, we focus on purchase prediction rather than CTR prediction; we aim to overcome the data sparsity of purchase records by leveraging their relationships with click records. Hence, the position bias in terms of click models is out of scope for our current work.
%\begin{table}[]
%	\centering
%	\small
%	\captionsetup{font=normal}
%	\caption{Notations.}
%	\vspace{-3ex}
%	\label{notation}
%	\begin{tabular}{l|l}
%		\specialrule{.1em}{.1em}{.1em} 
%		Symbol & \multicolumn{1}{c}{Description} \\ \specialrule{.1em}{.1em}{.1em}
%		$\mathcal{U}, \mathcal{I}$& Set of Users, Set of Items                                 \\ 		
%%		$n, m$& Number of users and items                                 \\ 
%		$\textbf{P}\in\mathbb{R}^{n\times m}$& User-Item Purchase matrix \\ 
%		$\textbf{B}\in\mathbb{R}^{n\times m}$& User-Item Click matrix \\ 
%		$\textbf{p}_u, \textbf{c}_u$& Items purchased, clicked by user $u$ \\ 
%%		$\textbf{c}_u$& Items clicked by user $u$ \\ 		
%		$\bm{\alpha \in \mathbb{R}}^{n\times K}$& User latent matrix                                 \\ 
%		$\bm{\beta \in \mathbb{R}}^{m\times K}$& Item latent matrix                                 \\ 
%		$\bm{\gamma \in \mathbb{R}}^{m}$& Item bias                                 \\ 
%%		$K$& Number of latent dimensions                                 \\ 
%		${\lambda}$& The strength of the model regularization                                 \\ 
%		$K,{\eta}$& Num. latent dimensions, Learning rate \\\specialrule{.1em}{.1em}{.1em}
%	\end{tabular}
%\vspace{-5ex}
%\end{table}

\medskip
\section{Conclusion \& Future Work}
In this paper, we introduced a framework that leverages users' past click records to complement the missing user--item interactions of purchase records. To this end, we formulated various model assumptions that define the order of user preferences regarding the non-purchased items, and demonstrated that click records are indeed useful for purchase prediction. 
We then proposed a novel learning-to-rank method,~\proposetwo, that is customized to be robust to relatively unreliable click records by particularly focusing on the accuracy of the top-ranked items.
%We then proposed a novel top-N focused learning-to-rank method that is deliberately designed to be robust to noisy click records, and at the same time focus on the accuracy near the top. 
We conducted extensive experiments on two real-world e-commerce datasets and verified the benefit of our proposed method compared with the state-of-the-art baselines.  We believe that our method is beneficial to any e-commerce stores, such as \textit{Amazon} and \textit{eBay}, which collect both purchase and click records.

For future work, we plan to extend our framework 1) to model the temporal and sequential information~\textcolor{black}{\cite{quadrana2018sequence}} of clicks and purchases by using Markov chain--based methods~\cite{rendle2010factorizing} or deep learning--based approaches such as recurrent neural networks~\cite{liu2016context,zhang2014sequential} and convolutional neural networks~\cite{liu2015convolutional}, 2) to incorporate side information related to users and items, such as user reviews and item images, to enhance the performance of the purchase prediction even further, 
\textcolor{black}{
and 3) to incorporate click count information for purchase prediction. Although the click counts are important when the candidate purchase items are clicked-but-not-purchased items, but not as important when the candidate purchase items are items neither clicked nor purchased (as in our setting). However, we think that it will be interesting to see how the click counts would help purchase prediction in our setting.
}

\section{Acknowledgment}
This research was supported by the Ministry of Science and ICT (MSIT), Korea,
under the Information and Communication Technology (ICT) Consilience Creative program
(IITP-2019-2011-1-00783) and (IITP-2018-0-00584) supervised by the Institute
for Information \& communications Technology Planning \& Evaluation (IITP), and Basic
Science Research Program through the National Research Foundation of Korea (NRF)
funded by the MSIT (NRF-2017M3C4A7063570) and (NRF-2016R1E1A1A01942642).
\nocite{Park:2015:PUP:2813448.2813517,Park:2016:TET:2872518.2889362,Park:2017:APH:3038912.3052581,Kim:2016:CMF:2959100.2959165}
\section*{References}
\bibliography{P3S_Journal}

\end{document}